\documentclass[letterpaper, 10 pt, conference]{ieeeconf}  

\usepackage[english]{babel} 
\usepackage[T1]{fontenc}
\usepackage[utf8x]{inputenc}
\usepackage{float}
\usepackage{graphicx}
\usepackage{amsfonts}
\usepackage{amssymb}
\usepackage{verbatim}
\usepackage{subfigure}
\usepackage{amsmath}
\usepackage{epstopdf}
\usepackage{fancyhdr}
\usepackage{booktabs,array}
\usepackage{soul}
\usepackage{color}
\usepackage{algpseudocode}
\usepackage{algorithm}
\usepackage{mathrsfs} 
\usepackage{siunitx}
\usepackage{hyperref}

\newtheorem{conj}{\textbf{Conjecture}}[section]

\IEEEoverridecommandlockouts                              

\title{\LARGE \bf
Optimal Time-Invariant Formation Tracking \\for a Second-Order Multi-Agent System
}

\author{Marco Fabris$^{1}$, Angelo Cenedese$^{1}$ and John Hauser$^{2}$ 
\thanks{$^{1}$M.Fabris and A.Cenedese are with the Department of Information Engineering,
	    University of Padova, via Gradenigo 6, 35131 Padova, Italy.
          E-mail: {\tt\small marco.fabris.7@phd.unipd.it}, 
                       {\tt\small angelo.cenedese@unipd.it}.}%
\thanks{$^{2}$J.Hauser is with the Department of Electrical, Computer \& Energy Engineering,
	    University of Colorado Boulder, 425 UCB \#1B55, Boulder, CO 80309, USA. 
        E-mail: {\tt\small John.Hauser@colorado.edu}.}
}

\begin{document}

\maketitle
\thispagestyle{empty}
\pagestyle{empty}

\begin{abstract}
Given a multi-agent linear system, we formalize and solve a trajectory optimization problem that encapsulates trajectory tracking, distance-based formation control and input energy minimization. To this end, a numerical projection operator Newton's method is developed to find a solution by the minimization of a cost functional able to capture all these different tasks. To stabilize the formation, a particular potential function has been designed, allowing to obtain specified geometrical configurations while the barycenter position and velocity of the system follows a desired trajectory.
\end{abstract}

\section{Introduction}
During the last decade, designing and controlling groups of robots to achieve specific collective goals has drawn a considerable interest. An individual agent may be programmed to be fully autonomous, but because of physical and resource constraints, its abilities may be limited. On the other hand, groups of individuals exchanging information and optimally self-organizing may have larger capabilities. 
In nature, examples of interacting \textquotedblleft swarms\textquotedblright$~$abound 
\cite{CamazineDeneubourgFranks2001}-\cite{ParrishEdelstein1999} 
and, such inspired, multi-agent systems have been widely designed to be used in applications like vehicle coordination, exploration and mapping of unknown environments, cooperative transportation, surveillance and monitoring of dynamic scenes and crowds. 

\subsubsection*{Related work} In the scientific community, a significant research effort has focused on the control of multi-agent systems and its mathematical foundation on graph-based networks~\cite{MesbahiEgerstedt2010}. 
Open challenges can be categorized as either formation control problems (with applications to mobile robots, autonomous underwater vehicles, satellites and spacecraft systems, automated highway and unmanned air vehicles, e.g. \cite{SchianoFranchiZelazo2016}-\cite{FathianDoucetteCurtis2018}), or other cooperative control problems such as role assignment, payload transport, air traffic control, timing, search and synchronization (e.g. \cite{CenedeseFavarettoOccioni2016}).\\
Formation Control generally aims to drive multiple agents to achieve prescribed constraints on their states. Depending on the sensing and the interaction capabilities of the agents, a variety of formation control problems can be found in the literature (see \cite{AndersonYuFidan2008} for an excellent overview on the subject). 
Nonetheless, creation and maintenance of a formation needs to trade off with single dynamics behaviors, collision avoidance, and dispersion~\cite{HauserCook2016}. In \cite{Reynolds1987}, an agent model that implements simple laws to abide these targets is introduced, thus determining the behavior of each agent within the group. These and similar laws often resort to a potential that determines the pairwise interactions among the agents, and along this line, many formation control algorithms adopt a formulation that contains an attractive part to maintain the swarm cohesion and a repulsive one to avoid agent collision (as the Morse potential in \cite{DOrsognaChuangBertozzi2006}-\cite{ChuangHuangDOrsogna2007}).
More generally, authors in \cite{OhParkAhn2015} discuss results on Formation Control discerning among position-based, displacement-based and distance-based according to the types of sensed and controlled variables. 
In particular, the last approach is that adopted in this work and can be defined in a framework where inter-agent distances are actively controlled to achieve the desired formation, which is defined through the specific inter-agent distances. Finally, in this respect, it is worth to highlight that several novel achievements have been developing by using distance-based control laws and an extensive part of that research has been dedicated to Formation Tracking \cite{SalinasBricarie2017}-\cite{YangCaoGarcia2018}. 

\subsubsection*{Contributions and outline of the paper} Many recent works developed for autonomous vehicles exploit Trajectory Optimization to perform maneuver regulation and motion planning, even in constrained environments. Practically, the employment of direct methods for the minimization of a cost functional represent a fundamental approach to provide solutions to this kind of problems: numerical tools, such as the PRojection Operator based Newton's method for Trajectory Optimization (PRONTO), have been successfully implemented in several instances \cite{BayerHauser2012}-\cite{AguiarPedroBayer2017}. At the light of this consideration, we propose here a novel application of PRONTO that aims at computing a solution for an optimization problem combining aspects of trajectory tracking, formation control and energy minimization.

In the remainder of the paper, Sections \ref{sec:problem_setup} and \ref{sec:numerical_methodologies_for_OIFT} describe how the specific problem of Formation Tracking can be formulated and solved by using PRONTO. Then, Sections \ref{sec:numerical_simulations} and \ref{sec:conclusions} show some numerical simulations, providing a validation of the proposed approach, interesting remarks and future directions for this research.

\section{Problem setup}\label{sec:problem_setup}
In this section, we discuss the assumptions regarding the agents' dynamics and how these lead to a simplified analysis, formalization and implementation of a centralized algorithm to solve an optimal control problem that encapsulates distinct tasks such as trajectory tracking, time-invariant formation control and input energy minimization. 
In particular, we address this as the Optimal Time-Invariant Formation Tracking (OIFT) problem, with the aim of finding a potential-based solution by minimizing a global cost functional able to capture all the different assignments simultaneously. Besides, it is crucial to highlight that communication topology constraints are not taken into account, since this paper lays the groundwork for a new optimization approach based on PRONTO in the field of Formation Tracking.

\subsection{Agents dynamics}
We assume that $n>1$ robotic agents have been already deployed on a $M$-dimensional space, with $M\in \left\lbrace 1,2,3\right\rbrace  $. We also suppose that each agent $i$, for $i=1,...,n$, is aware of its absolute position $\mathbf{p}_{i} \in \mathbb{R}^{M}$ and velocity $\dot{\mathbf{p}}_{i}\in \mathbb{R}^{M}$ and can be driven by regulating its absolute acceleration $\ddot{\mathbf{p}}_{i}\in \mathbb{R}^{M}$. For the sake of simplicity, a linear dynamics for the robots is adopted. 
With $N:=nM$, the expressions of the state $\mathbf{x}\in \mathbb{R}^{2N}$ and the input $\mathbf{u}\in \mathbb{R}^{N}$ of this linear system of mobile elements are given respectively by
\begin{align*}
\mathbf{x} &= \begin{bmatrix}
\mathbf{p}_{1}^{\top} & \dots & \mathbf{p}_{n}^{\top} & \dot{\mathbf{p}}_{1}^{\top} & \dots & \dot{\mathbf{p}}_{n}^{\top} 
\end{bmatrix}^{\top} &&= \begin{bmatrix}
\mathbf{p}^{\top} & \dot{\mathbf{p}}^{\top}
\end{bmatrix}^{\top};\\
\mathbf{u} &= 
\begin{bmatrix}
\ddot{\mathbf{p}}_{1}^{\top} & \dots & \ddot{\mathbf{p}}_{n}^{\top}
\end{bmatrix}^{\top} &&= \ddot{\mathbf{p}}.
\end{align*}
We finally assume that the state information is globally available to all  agents, so that an estimate of the barycenter position $\mathbf{p}_{B}\!=\!\dfrac{1}{n} \sum\limits_{i=1}^{n} \mathbf{p}_{i}$ and velocity $\dot{\mathbf{p}}_{B}$ be available to each agent at each time instant. We set
$\mathbf{x}_{B} = \begin{bmatrix}
\mathbf{p}_{B}^{\top} & \dot{\mathbf{p}}_{B}^{\top}
\end{bmatrix}^{\top}$,
with $\mathbf{x}_{B}\in \mathbb{R}^{2M}$, to be an output for this linear system.
Since we wish to govern positions and velocities of the agents by controlling their accelerations, a double integrator model is a suitable choice for this purpose: the second-order dynamics that follows can be represented via the linear state space
\begin{equation}\label{eq:dyn_sys}
\begin{cases}
\dot{\mathbf{x}} = \mathbf{A} \mathbf{x} + \mathbf{B} \mathbf{u}\\
\mathbf{x}_{B} = \mathbf{C}\mathbf{x}
\end{cases}.
\end{equation}
State matrix $\mathbf{A}\in \mathbb{R}^{2N\times 2N}$, input matrix $\mathbf{B}\in \mathbb{R}^{2N\times N}$ and output matrix $\mathbf{C} \in \mathbb{R}^{2M\times 2N}$ in \eqref{eq:dyn_sys} are given by
%
%
\begin{equation*}
\mathbf{A}=\begin{bmatrix}
\mathbf{Z}_{N} & \mathbf{I}_{N}\\
\mathbf{Z}_{N} & \mathbf{Z}_{N}
\end{bmatrix}, 
\qquad
\mathbf{B} = \begin{bmatrix}
\mathbf{Z}_{N}\\
\mathbf{I}_{N}
\end{bmatrix}
\end{equation*}
\begin{equation*}
\mathbf{C} = \dfrac{1}{n} \begin{bmatrix}
\mathbf{I}_{M} & \dots & \mathbf{I}_{M} & \mathbf{Z}_{M} & \dots & \mathbf{Z}_{M}\\
\mathbf{Z}_{M}  & \dots &  \mathbf{Z}_{M} & \mathbf{I}_{M} & \dots & \mathbf{I}_{M} 
\end{bmatrix},
\end{equation*}
where $\mathbf{I}_{\natural}$ denotes the $\natural$-dimensional identity matrix, $\mathbf{Z}_{\natural_{1}\times \natural_{2}}$ indicates null matrices of dimension $\natural_{1}\!\times\!\natural_{2}$ and if $\natural_{1}\!=\!\natural_{2}\!=\!\natural$ the convention $\mathbf{Z}_{\natural\times \natural}=\mathbf{Z}_{\natural}$ is adopted instead.


\subsection{Cost functional minimization}
Tracking a desired path 
$\mathbf{x}_{B,des}=\begin{bmatrix}
\mathbf{p}_{B,des}^{\top} & \dot{\mathbf{p}}_{B,des}^{\top}
\end{bmatrix}^{\top}$ 
with the centroid $\mathbf{x}_{B}$ while minimizing the energy spent by the input $\mathbf{u}$ and attaining a specified formation during this optimal control procedure is what we aim at.\\
Let $\mathscr{T}$ be the trajectory manifold of \eqref{eq:dyn_sys}, such that $\boldsymbol{\xi} = \left(\mathbf{x}(\cdot),\mathbf{u}(\cdot)\right) \in \mathscr{T}$. The general optimization problem can be formulated as follows: find a solution $\boldsymbol{\xi}^{\star}$ such that
\begin{equation*}
\underset{\boldsymbol{\xi}\in \mathscr{T}}{\min}~ h(\boldsymbol{\xi})
\end{equation*}
is attained, where
\begin{equation}\label{eq:cost}
h(\mathbf{x}(\cdot),\mathbf{u}(\cdot)) = \int_{0}^{T} l\left(\mathbf{x}(\tau), \mathbf{u}(\tau), \tau \right) d\tau + m(\mathbf{x}(T))
\end{equation}
represents the cost functional to be minimized in order to fulfill the three objectives previously stated. In \eqref{eq:cost}, two different terms explicitly appear:
the instantaneous cost
\begin{equation}\label{eq:inst_cost}
l\left(\mathbf{x}(\tau), \mathbf{u}(\tau), \tau \right) := l^{tr}(\mathbf{x}_{B}) + l^{in}(\mathbf{u}) + l_{d}^{fo}(\mathbf{p})
\end{equation}
and the final cost
\begin{equation}\label{eq:final_cost_m}
m(\mathbf{x}(T)) := \mathbf{x}^{\top}(T) \mathbf{P}_{1} \mathbf{x}(T)
\end{equation}
in which, matrix $\mathbf{P}_{1}$ is set to be null in order to focus our analysis on the instantaneus cost \eqref{eq:inst_cost} and simplify the overall framework.
Each term in \eqref{eq:inst_cost} is minimized with the aim to obtain the achievement of a specific task. Indeed, setting $r_{ij}=\left\|\mathbf{p}_{i}-\mathbf{p}_{j}\right\|_{2}$ as the inter-agent Euclidean distance, we can define each contribution as follows:
\begin{equation}\label{eq:track_inst_cost}
l^{tr}(\mathbf{x}_{B}) = \dfrac{1}{2} \left\|\mathbf{x}_{B}-\mathbf{x}_{B,des}\right\|_{\mathbf{Q}_{B}}^{2}
\end{equation}
for the tracking task,
\begin{equation*}
l^{in}(\mathbf{u}) = \dfrac{1}{2} \left\|\mathbf{u}\right\|_{\mathbf{R}}^{2}
\end{equation*}
for the input energy task and, given a family of potential functions $\sigma_{d_{ij}}:\mathbb{R}_{\geq 0} \rightarrow \mathbb{R}_{\geq 0}$,
\begin{equation}\label{eq:form_inst_cost}
l_{d}^{fo}(\mathbf{p}) = \dfrac{k_{F}}{2} \sum\limits_{i=1}^{n} \sum\limits_{\forall j\neq i} \sigma_{d_{ij}}\left(r^{2}_{ij}\right)
\end{equation}
for the formation task.\\
Symmetric positive semidefinite matrix $\mathbf{Q}_{B}\in \mathbb{R}^{2M\times 2M}$, symmetric positive definite matrix $\mathbf{R}\in \mathbb{R}^{N\times N}$ and $k_{F}>0$ are constant weights that can be tuned according to given specifications in order to penalize the trajectory tracking, the energy spent by the inputs and the convergence to a desired shape respectively.
Furthermore, with regard to potential functions $\sigma_{d_{ij}} $ in formation cost \eqref{eq:form_inst_cost}, each $d_{ij}$ represents the desired inter-agent distance between the pair of agents $(i,j)$: an accurate choice of $\sigma_{d_{ij}}$ allows the usage of the optimization framework PRONTO, as explained next. 
 	
\subsection{Potential based formations}
As for the formation objective, we would like the system of agents to achieve desired linear, 2D or 3D shapes induced by a set of constraints of the form
\begin{equation}\label{eq:dist_constraints}
r_{ij} = d_{ij}, \quad \text{s.t. } i=1,...,n, ~\forall j\neq i;
\end{equation}
therefore, potential functions to accomplish this task are introduced. Let $s_{ij}=r^{2}_{ij}$ be the squared inter-agent distance between agents $i$ and $j$. Among various possible choices, the potential function
\begin{equation}\label{eq:potential}
\sigma_{d_{ij}}(s_{ij}) := \begin{cases}
k_{r} (1-s_{ij}/d_{ij}^{2})^{3} \quad & \text{for } 0\leq s_{ij} \leq d_{ij}^{2}  \\
k_{a} \left(\sqrt{s_{ij}}/d_{ij}-1\right)^{3} \quad& \text{for } s_{ij} \geq d_{ij}^{2}
\end{cases}
\end{equation}
has been selected since it is polynomial in $r_{ij}$ with relatively low degree and differentiable with respect to $s_{ij}\geq 0$ until the second order. In particular, \eqref{eq:potential} is one of the $\mathscr{C}^{2}$ functions with lowest polynomial order that can be adjusted by tuning constants $k_{r}>0$ and $k_{a}>0$ independently. These two hyper-parameters are purposely designed to be directly proportional to the repulsion and attraction action between agents respectively, playing a key role in the intensity regulation of the potential itself (see Fig. \ref{fig:sigma} in Sec. \ref{ssec:PRONTO_heuristics}).

The use of functions with the same properties of \eqref{eq:potential} can be justified by the fact that they can lead formations to satisfy the maximum number of feasible\footnote{I.e., that can be satisfied concurrently.} relations in \eqref{eq:dist_constraints}, since the dynamics is purposely steered to minimum-potential trajectories. Indeed, it can be easily proven that $\sigma_{d_{ij}}(s_{ij})$ is nonnegative for all $s_{ij}$ and exhibits a unique global minimum point $\left(s_{ij},\sigma_{d_{ij}}(s_{ij})\right)=\left(d^{2}_{ij},0\right)$, meaning that this potential vanishes when the desired distance $d_{ij}$ between the pair $(i,j)$ is attained.

\section{Numerical methodologies for OIFT} \label{sec:numerical_methodologies_for_OIFT}
We provide here a brief introduction to PRONTO\footnote{For further details, the interested reader is referred to the accurate explanations provided in chapter 4 of thesis \cite{Hausler2015} and in \cite{HauserSaccon2006}.}, the numerical tool used for trajectory optimization, and  clarify its adaptation to solve the OIFT problem introduced before. 

\subsection{Basics of PRONTO}\label{subs:basics_of_PRONTO}
In its simplest form, PRONTO  is an iterative numerical algorithm for the minimization, via a Newton method, of a general cost functional \eqref{eq:cost}. To work on a trajectory manifold $\mathscr{T}$, one projects state-control curves in the ambient Banach space onto $\mathscr{T}$ by using a linear time-varying trajectory tracking controller. 
To this end, a nonlinear feedback system
\begin{align*}
\dot{\mathbf{x}} &= f(\mathbf{x},\mathbf{u}), \quad \mathbf{x}(0) = \mathbf{x}_{0}; \\
\mathbf{u}       &= \boldsymbol{\mu} + \mathbf{K} \left(\boldsymbol{\alpha}-\mathbf{x}\right), 
\end{align*}
where matrix $\mathbf{K}$ acts as a time-varying controller, defines a nonlinear operator ($\mathscr{C}^{r}$ when $f\in \mathscr{C}^{r}$)
\begin{equation*}\label{eq:PRONTO_def}
\mathscr{P}\left(\boldsymbol{\xi}\right) : \boldsymbol{\xi}=\left(\boldsymbol{\alpha}(\cdot),\boldsymbol{\mu}(\cdot)\right) \mapsto \boldsymbol{\eta} = \left(\mathbf{x}(\cdot),\mathbf{u}(\cdot)\right)
\end{equation*}
mapping bounded curves (in its domain) to trajectories.
 
Among the many properties of $\mathscr{P}(\cdot)$, the most relevant in terms of the practical use of PRONTO refers to the fact that it makes possible to easily switch between constrained and unconstrained minimization to solve the initial problem, as
\begin{equation*}
\underset{\boldsymbol{\xi} \in \mathscr{T}}{\min}~h(\boldsymbol{\xi}) \quad \Leftrightarrow \quad \underset{\boldsymbol{\xi}}{\min}~g(\boldsymbol{\xi})
\end{equation*}
where $g(\boldsymbol{\xi}) = h\left(\mathscr{P}(\boldsymbol{\xi})\right)$.

Let us denote the Fréchet derivative of $\mathscr{P}(\cdot)$ with the continuous linear operator
\begin{equation*}
D\mathscr{P}(\boldsymbol{\xi}) : \boldsymbol{\zeta} = \left(\boldsymbol{\beta}(\cdot),\boldsymbol{\nu}(\cdot)\right) \mapsto \boldsymbol{\gamma} = \left(\mathbf{z}(\cdot), \mathbf{v}(\cdot)\right) 
\end{equation*}
such that the approximation 
\begin{equation*}
\mathscr{P}(\boldsymbol{\xi}+\boldsymbol{\zeta}) \approx \mathscr{P}(\boldsymbol{\xi}) + D \mathscr{P}(\boldsymbol{\xi}) \cdot \boldsymbol{\zeta}
\end{equation*}
holds for all $\boldsymbol{\xi} \in \mathscr{T}$. Then, PRONTO algorithm can be illustrated as follows:
\vspace{-6pt}
\newpage$~$\\ \vspace{-1.1cm}
\begin{algorithm}
	\caption{Projection operator Newton's method \cite{Hauser2002}}\label{alg:PRONTO}
	\begin{algorithmic}[1]
		\State 		\textbf{given} initial trajectory $\boldsymbol{\xi}_{0} \in \mathscr{T}$
		\For{$k = 0, 1, 2, ...$} 
		\State Redesign feedback $\mathbf{K}(\cdot)$ for $\mathscr{P}(\cdot)$, if desired/needed \label{redesign_controller}
		\vspace{-0.3cm}
		\State $\boldsymbol{\zeta}_{k} = \underset{\boldsymbol{\zeta} \in T_{\boldsymbol{\xi}_{k}} \mathscr{T}}{\arg \min}~Dh\left(\boldsymbol{\xi}_{k}\right) \cdot \boldsymbol{\zeta}  +  \dfrac{1}{2} D^{2}g\left(\boldsymbol{\xi}_{k}\right) \cdot \left(\boldsymbol{\zeta}, \boldsymbol{\zeta}\right)$ \label{search_direction}
		\State $\gamma_{k} = \underset{\gamma \in (0,1]}{\arg \min} ~ g\left(\boldsymbol{\xi}_{k}+\gamma \boldsymbol{\zeta}_{k}\right)$ \label{step_size}
		\State $\boldsymbol{\xi}_{k+1} = \mathscr{P}\left(\boldsymbol{\xi}_{k}+\gamma_{k} \boldsymbol{\zeta}_{k}\right)$ \label{update}
	    \EndFor
	\end{algorithmic}
\end{algorithm}
\vspace{-6pt}

\noindent Step \ref{search_direction} in Alg. \ref{alg:PRONTO} encodes the search direction problem and can be solved by finding a solution to the time-varying linear quadratic (LQ) optimal control problem of the form
\begin{align}
	&\underset{}{\min} \int_{0}^{T} \left(\mathbf{a}^{\top}\mathbf{z}+\mathbf{b}^{\top}\mathbf{v}+\dfrac{1}{2} \begin{bmatrix}
	\mathbf{z} \\ \mathbf{v}
	\end{bmatrix}^{\top}  \begin{bmatrix}
	\mathbf{Q}_{o} 		  & \mathbf{S}_{o} \\
	\mathbf{S}_{o}^{\top} & \mathbf{R}_{o}
	\end{bmatrix} \begin{bmatrix}
	\mathbf{z} \\ \mathbf{v}
	\end{bmatrix}\right) d\tau +\nonumber \\ 
	&\quad\quad\quad\quad\quad\quad\quad\quad\quad\quad\quad\quad \mathbf{r}_{1}^{\top}\mathbf{z}(T) + \dfrac{1}{2} \mathbf{z}(T)^{\top} \mathbf{P}_{1}\mathbf{z}(T) \nonumber\\
	&\text{subject to } \dot{\mathbf{z}} = \bar{\mathbf{A}} \mathbf{z} + \bar{\mathbf{B}} \mathbf{v}, \quad \mathbf{z}(0) = \mathbf{z}_{0} \label{eq:line_search}
\end{align}
where vectors $\mathbf{a}$, $\mathbf{b}$ and matrices $\mathbf{Q}_{o}$, $\mathbf{S}_{o}$, $\mathbf{R}_{o}$, $\bar{\mathbf{A}}$, $\bar{\mathbf{B}}$ are time-varying parameters descending from the instantaneous cost $l$ and the dynamics adopted. Moreover, vector $\mathbf{r}_{1}$ and matrix $ \mathbf{P}_{1}$ are given constants depending on the final cost \eqref{eq:final_cost_m}.
When problem \eqref{eq:line_search} possesses a unique minimizing trajectory, it can be solved resorting to the following differential Riccati equation in the matrix variable $\mathbf{P}$:
\begin{equation}\label{eq:Riccati_for_search_dir}
\begin{cases}
-\dot{\mathbf{P}} \!=\! \bar{\mathbf{A}}^{\top} \mathbf{P} \!+\! \mathbf{P} \bar{\mathbf{A}} \!-\! \mathbf{K}_{o}^{\top}\mathbf{R}_{o} \mathbf{K}_{o} \!+\! \mathbf{Q}_{o}, \quad\! \mathbf{P}(T)\!=\! \mathbf{P}_{1}\\
\mathbf{K}_{o} \!=\! \mathbf{R}_{o}^{-1} \left(\mathbf{S}_{o}^{\top} \!+\! \bar{\mathbf{B}}^{\top} \mathbf{P}  \right)
\end{cases}
\end{equation}

\subsection{Application of PRONTO to the OIFT problem}
As previously mentioned, the assumptions made in Sec. \ref{sec:problem_setup} allow to obtain significant simplifications in the expression of many numerical quantities while preserving the possibility to perform interesting trajectory explorations for the system of agents. Now, we clarify how PRONTO variables and parameters declared in Alg. \ref{alg:PRONTO} have to be practically assigned and computed when this particular framework is adopted. Indeed, setting
$f(\mathbf{x},\mathbf{u}) = \mathbf{A} \mathbf{x} + \mathbf{B} \mathbf{u}$,
and denoting the zero vector of dimension $\natural$ with $\mathbf{0}_{\natural}$, the following list of equalities is immediately available:  
\begin{alignat}{3}
& \bar{\mathbf{A}} &&= f_{\mathbf{x}}           &&= \mathbf{A} \nonumber\\
& \bar{\mathbf{B}} &&= f_{\mathbf{u}}           &&= \mathbf{B} \nonumber\\
& \mathbf{a}       &&= l^{\top}_{\mathbf{x}}    &&= \mathbf{C}^{\top}\mathbf{Q}_{B}\left(\mathbf{x}_{B}-\mathbf{x}_{B,des}\right) + \nabla_{\mathbf{x}}l_{d}^{fo} \label{eq:a}\\
& \mathbf{b}       &&= l^{\top}_{\mathbf{u}}    &&= \mathbf{R} \mathbf{u} \nonumber\\
& \mathbf{r}_{1}   &&= m^{\top}_{\mathbf{x}}    &&= \mathbf{0}_{2N} \nonumber\\
& \mathbf{Q}_{o}   &&= l_{\mathbf{x}\mathbf{x}} &&= \mathbf{C}^{\top}\mathbf{Q}_{B}\mathbf{C} + \mathcal{H}_{\mathbf{x}\mathbf{x}} l_{d}^{fo} \label{eq:Qo}\\
& \mathbf{S}_{o}   &&= l_{\mathbf{x}\mathbf{u}} &&= \mathbf{Z}_{2N\times N} \nonumber\\
& \mathbf{R}_{o}   &&= l_{\mathbf{u}\mathbf{u}} &&= \mathbf{R} \nonumber\\
& \mathbf{P}_{1}   &&= m_{\mathbf{x}\mathbf{x}} &&= \mathbf{Z}_{2N}\nonumber .
\end{alignat}
where symbols $\nabla_{*}$ and $\mathcal{H}_{**}$ indicate standard gradient and Hessian operators.\\
Hence, a constant proportional derivative (PD) controller as
\begin{equation}\label{eq:PDcontroller}
\mathbf{K} = \begin{bmatrix}
k_{p} \mathbf{I}_{N} & k_{v} \mathbf{I}_{N}
\end{bmatrix}, \quad k_{p}, k_{v} > 0
\end{equation}
is the simplest and most efficient choice to opt for
, meaning that step \ref{redesign_controller} in Alg. \ref{alg:PRONTO} is not needed to be processed. In addition, let us examine the gradient and Hessian matrix of $l_{d}^{fo}$ w.r.t. the state $\mathbf{x}$ appearing in \eqref{eq:a} and \eqref{eq:Qo} respectively. Their formal expressions are yielded by
\begin{alignat*}{3}
&\nabla_{\mathbf{x}}l_{d}^{fo} &&= \begin{bmatrix}
\nabla^{\top}_{\mathbf{p}}l_{d}^{fo} & \nabla^{\top}_{\dot{\mathbf{p}}}l_{d}^{fo}
\end{bmatrix}^{\top} &&= \begin{bmatrix}
\nabla^{\top}_{\mathbf{p}}l_{d}^{fo} & \mathbf{0}^{\top}_{N}
\end{bmatrix}^{\top}\\
&\mathcal{H}_{\mathbf{x}\mathbf{x}} l_{d}^{fo} &&= \begin{bmatrix}
\mathcal{H}_{\mathbf{p}\mathbf{p}} l_{d}^{fo} & \mathcal{H}_{\mathbf{p}\dot{\mathbf{p}}} l_{d}^{fo} \\
\mathcal{H}_{\dot{\mathbf{p}}\mathbf{p}} l_{d}^{fo} & \mathcal{H}_{\dot{\mathbf{p}}\dot{\mathbf{p}}} l_{d}^{fo}
\end{bmatrix} &&= \begin{bmatrix}
\mathcal{H}_{\mathbf{p}\mathbf{p}} l_{d}^{fo} &  \mathbf{Z}_{N}  \\
\mathbf{Z}_{N}								  &  \mathbf{Z}_{N}
\end{bmatrix}
\end{alignat*}
where, by setting $\boldsymbol{\Pi}_{ij}=\left(\mathbf{p}_{i}-\mathbf{p}_{j}\right) \left(\mathbf{p}_{i}-\mathbf{p}_{j}\right)^{\top} \in \mathbb{R}^{M\times M}$, it holds that
\begin{equation*}
\nabla_{\mathbf{p}_{i}}l_{d}^{fo} = 2k_{F} \sum\limits_{\forall j\neq i} \sigma_{d_{ij}}^{\prime} \left(s_{ij}\right)  \left(\mathbf{p}_{i}-\mathbf{p}_{j}\right) 
\end{equation*}
and, for all $i\neq j$,
\begin{equation*}
\mathcal{H}_{\mathbf{p}_{i}\mathbf{p}_{j}} l_{d}^{fo} = -2k_{F}\left[2\sigma_{d_{ij}} ^{\prime\prime} \left(s_{ij}\right)  \boldsymbol{\Pi}_{ij} + \sigma_{d_{ij}}^{ \prime} \left(s_{ij}\right)  \mathbf{I}_{M} \right].
\end{equation*}
Note that the choice of having $\sigma_{d_{ij}} \in \mathscr{C}^{2}$ is required to use PRONTO. Furthermore, a nice expression for the diagonal blocks in the Hessian $\mathcal{H}_{\mathbf{p}\mathbf{p}} l_{d}^{fo}$ can be provided by
\begin{equation*}
\mathcal{H}_{\mathbf{p}_{i}\mathbf{p}_{i}} l_{d}^{fo} = - \sum\limits_{\forall j\neq i} \mathcal{H}_{\mathbf{p}_{i}\mathbf{p}_{j}} l_{d}^{fo}.
\end{equation*}

\subsection{Robust heuristic for PRONTO}\label{ssec:PRONTO_heuristics}
\begin{figure}[h!]
	\centering
	\includegraphics[width=0.35\textwidth, height=0.25\textwidth]{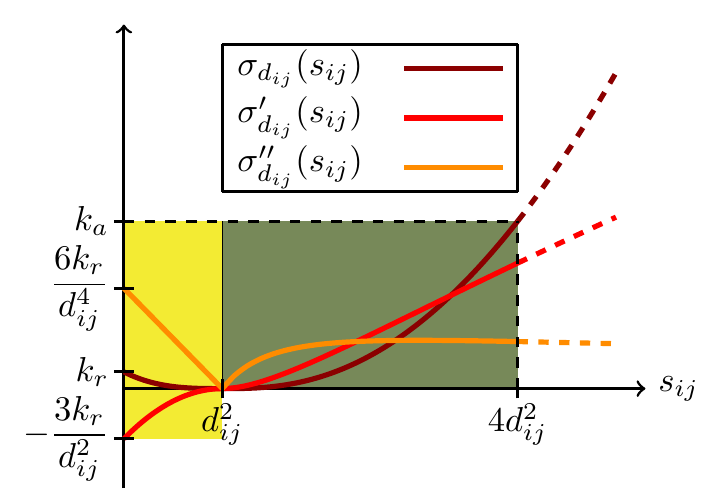}
	\caption{Potential function $\sigma_{d_{ij}}(s_{ij})$ with $k_{a}>k_{r}$ and its derivatives w.r.t. $s_{ij}$ up to the second order. Repulsive and attractive behaviors can be associated to the values in the yellow and dark green zones respectively.}
	\label{fig:sigma}
\end{figure}

As a matter of fact, the LQ problem in \eqref{eq:Riccati_for_search_dir} requires a positive semidefinite matrix $\mathbf{Q}_{o}$ to be solved; therefore, care has to be taken while search directions in Alg. \ref{alg:PRONTO} are computed. In reality, it can be noticed that expression \eqref{eq:Qo} does not always satisfy this condition because of the general undetermined definiteness of the Hessian term. To this purpose, we decide to implement a suitable safe version of $\mathbf{Q}_{o}$, namely $\mathbf{Q}^{safe}_{o}$, by using a Gershgorin-circle-theorem-based heuristic that can be justified by the fact that potential functions with the same repulsive-attractive behavior of \eqref{eq:potential} never guarantee the positive semidefiniteness of matrix $\mathcal{H}_{\mathbf{p}\mathbf{p}} l_{d}^{fo}$. Indeed, the latter term is badly affected by the evidence that matrix $\boldsymbol{\Pi}_{ij}$ is rank-$1$ for all the pairs $(i,j)$ and the possibility of $\sigma_{d_{ij}}^{\prime}(s_{ij})$ to be negative.\\
As illustrated in Fig. \ref{fig:sigma}, the qualitative graphic of $\eqref{eq:potential}$ and its derivatives highlights that
\begin{enumerate}
	\item $\sigma_{d_{ij}}^{\prime} (s_{ij}) \leq 0$ for $0\leq s_{ij}\leq d^{2}_{ij}$; \label{prop:sig_prime}
	\item $\sigma_{d_{ij}}^{\prime} (s_{ij}) \geq 0$ for $s_{ij}\geq d^{2}_{ij}$;
	\item $\sigma_{d_{ij}}^{\prime\prime} (s_{ij}) \geq 0$ for all $s_{ij}$; \label{prop:sig_prime_prime}
\end{enumerate}
where the equalities in \ref{prop:sig_prime})-\ref{prop:sig_prime_prime}) hold if and only if $s_{ij}=d^{2}_{ij}$.

The previous observations allow to state that, whenever two agents $(i,j)$ are repelling each other, the Hessian $\mathcal{H}_{\mathbf{p}\mathbf{p}} l_{d}^{fo}$ is not positive semidefinite. This implication leads us to neglect the variation of potential \eqref{eq:potential} w.r.t. the current distance for $0\leq s_{ij}<d^{2}_{ij}$ to aim at searching a well defined descent direction in Alg. \ref{alg:PRONTO}.
To this end, we finally choose to force the term $\mathcal{H}_{\mathbf{p}\mathbf{p}} l_{d}^{fo}$ to be positive semidefinite while its value is being computed, as implemented at lines \ref{if:Qsafe}-\ref{endif:Qsafe} of Alg. \ref{alg:Qsafe}.

\begin{algorithm}
	\caption{Heuristic for the modification of $\mathcal{H}_{\mathbf{p}\mathbf{p}} l_{d}^{fo}$ to search effectively a descent direction in PRONTO}\label{alg:Qsafe}
	\begin{algorithmic}[1]
		\For {$i = 1,...,n$}
		\State $\mathcal{H}_{\mathbf{p}_{i}\mathbf{p}_{i}} l_{d}^{fo} \leftarrow \mathbf{Z}_{M} $
		\For {$j = 1,...,n$ s.t. $j\neq i$}
		\State  $\mathcal{H}_{\mathbf{p}_{i}\mathbf{p}_{j}} l_{d}^{fo} \leftarrow -4k_{F}\sigma_{d_{ij}} ^{\prime\prime} \left(s_{ij}\right)  \boldsymbol{\Pi}_{ij} $
		\If {$\mathbf{Q}^{safe}_{o}$ is not required \textbf{or} $\sigma_{d_{ij}} ^{\prime} \left(s_{ij}\right) > 0$} \label{if:Qsafe}
		\State $\mathcal{H}_{\mathbf{p}_{i}\mathbf{p}_{j}} l_{d}^{fo} \leftarrow \mathcal{H}_{\mathbf{p}_{i}\mathbf{p}_{j}} l_{d}^{fo} -2k_{F} \sigma_{d_{ij}} ^{\prime} \left(s_{ij}\right) \mathbf{I}_{M} $
		\EndIf \label{endif:Qsafe}
		\State $\mathcal{H}_{\mathbf{p}_{i}\mathbf{p}_{i}} l_{d}^{fo} \leftarrow \mathcal{H}_{\mathbf{p}_{i}\mathbf{p}_{i}} l_{d}^{fo}-\mathcal{H}_{\mathbf{p}_{i}\mathbf{p}_{j}} l_{d}^{fo}$
		\EndFor
		\EndFor
	\end{algorithmic}
\end{algorithm}

\section{Numerical simulations} \label{sec:numerical_simulations}
In this work, we have mainly focused on the case where
\begin{equation}\label{eq:constraints_d_equal}
r_{ij} = d, \quad  i=1,...,n, ~\forall j\neq i
\end{equation}
is imposed, which, in most of the actual scenarios, becomes an infeasible\footnote{Feasibility trivially holds only for $n\leq M+1$.} requirement to be completely verified for each pair $(i,j)$. Set of constraints in \eqref{eq:constraints_d_equal} allows to explore specific formations, called $d$-configurations\footnote{That is the final configurations achieved by agents satisfying the maximum number of constraints in set \eqref{eq:constraints_d_equal}.}, where potentials can be interestingly tested. Thus, if not differently specified, we adopt \eqref{eq:constraints_d_equal} as an assumption with $d = 5~\si{\meter}$ during the dissertation. Whenever \eqref{eq:constraints_d_equal} becomes unfeasible, we relax the constraint to
\begin{equation}\label{eq:constraints_d_equal_relaxation}
\left|r_{ij}-d\right|/d < 10^{-1}.
\end{equation}
We have also chosen to keep the input weighting matrix $\mathbf{R} = r_{a}\mathbf{I}_{N}$, with $r_{a} = 1 ~\si{\meter^{-2} \second^{4}}$, and the integration time $T=20 ~\si{\second}$ fixed, as well as the structure of the output weighting matrix
\begin{equation*}
\mathbf{Q}_{B}=\begin{bmatrix}
q_{p} \mathbf{I}_{M} & \mathbf{Z}_{M} \\
\mathbf{Z}_{M} & q_{v}\mathbf{I}_{M}
\end{bmatrix},
\end{equation*}
where $q_{p}$, $q_{v}$ are nonnegative weights.
Moreover, we have set the following parameters: $k_{r} = 10^{2}$, $k_{a} = 1$, in \eqref{eq:potential} and $k_{F} = 10^{-1}$, in \eqref{eq:form_inst_cost}. In each numerical simulation presented, we have established to stop the execution of Alg. \ref{alg:PRONTO} by taking two actions: an interruption occurs when a preset maximum number of iterations $MaxIter > 0$ is exceeded; the algorithm can terminate at the iteration $k \leq MaxIter$ if, given a threshold $\varepsilon > 0$, the inequality $-Dg\left(\boldsymbol{\xi}_{k}\right) \cdot \boldsymbol{\zeta}_{k} < \varepsilon$ is satisfied. In particular, we have set $\varepsilon = 10^{-8}$.

Finally, constant gains for the PD controller in \eqref{eq:PDcontroller} have been assigned exploiting the linear dynamics of \eqref{eq:dyn_sys}, as
\begin{equation*}
\begin{cases}
\omega_{n} = 3 ~\si{\radian\per\second}\\
\xi = 0.7 
\end{cases} \Rightarrow ~ \begin{cases}
k_{p} = \omega_{n}^{2} \\
k_{v} = 2\xi\omega_{n}
\end{cases}.
\end{equation*}

\subsection{Validity test for the developed algorithm}
\begin{figure}[h!]
	\centering
	\hspace{-0.2cm}
	\subfigure[Line tracking ($M=2$, $n=3$)]{\includegraphics[height=3cm,width=4.2cm]{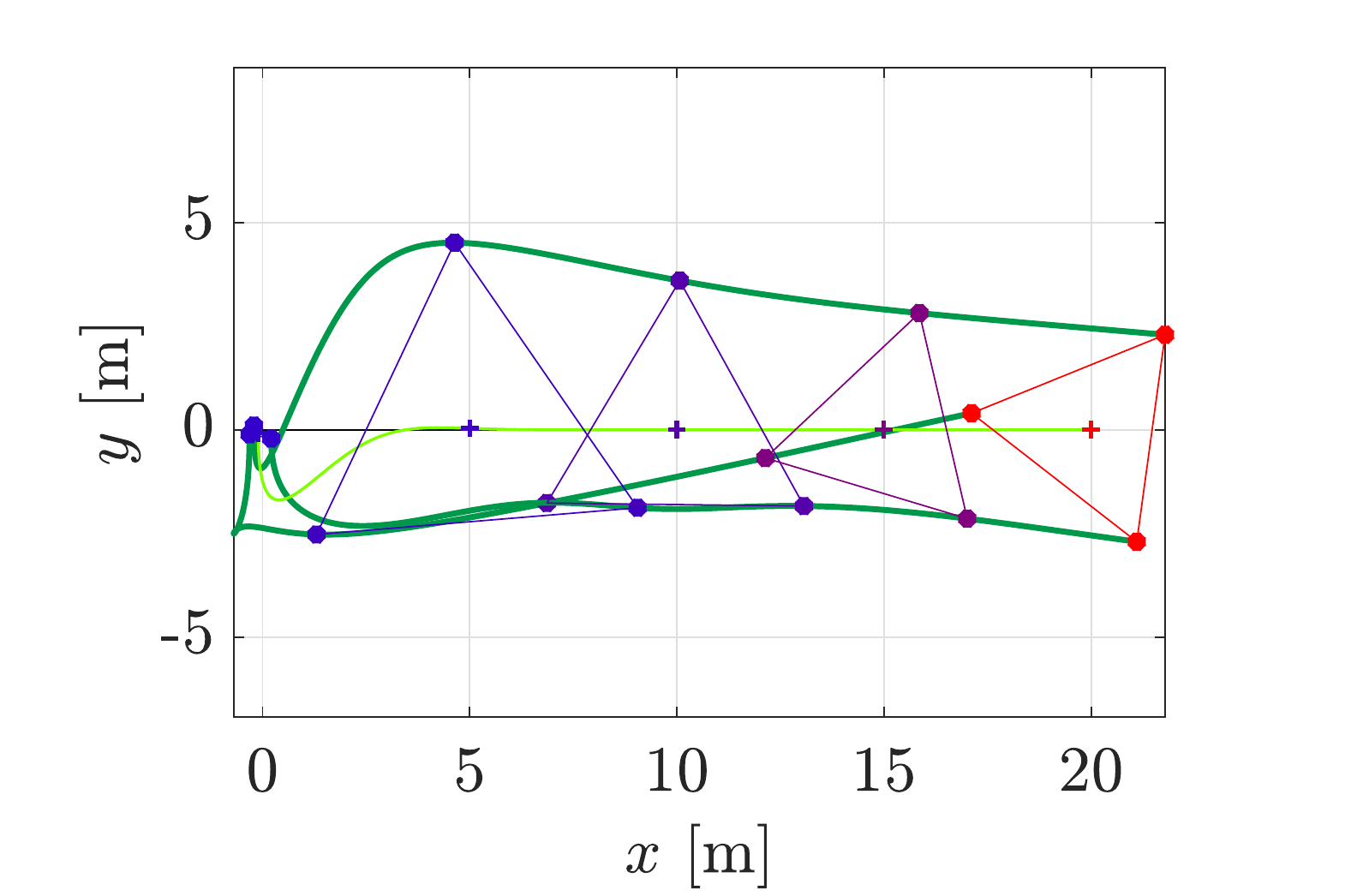}\label{fig:valid_2D_scene}} 
	\hspace{-0.19cm} 
	\subfigure[Cost function in (a)]{\includegraphics[height=3cm,width=4.2cm]{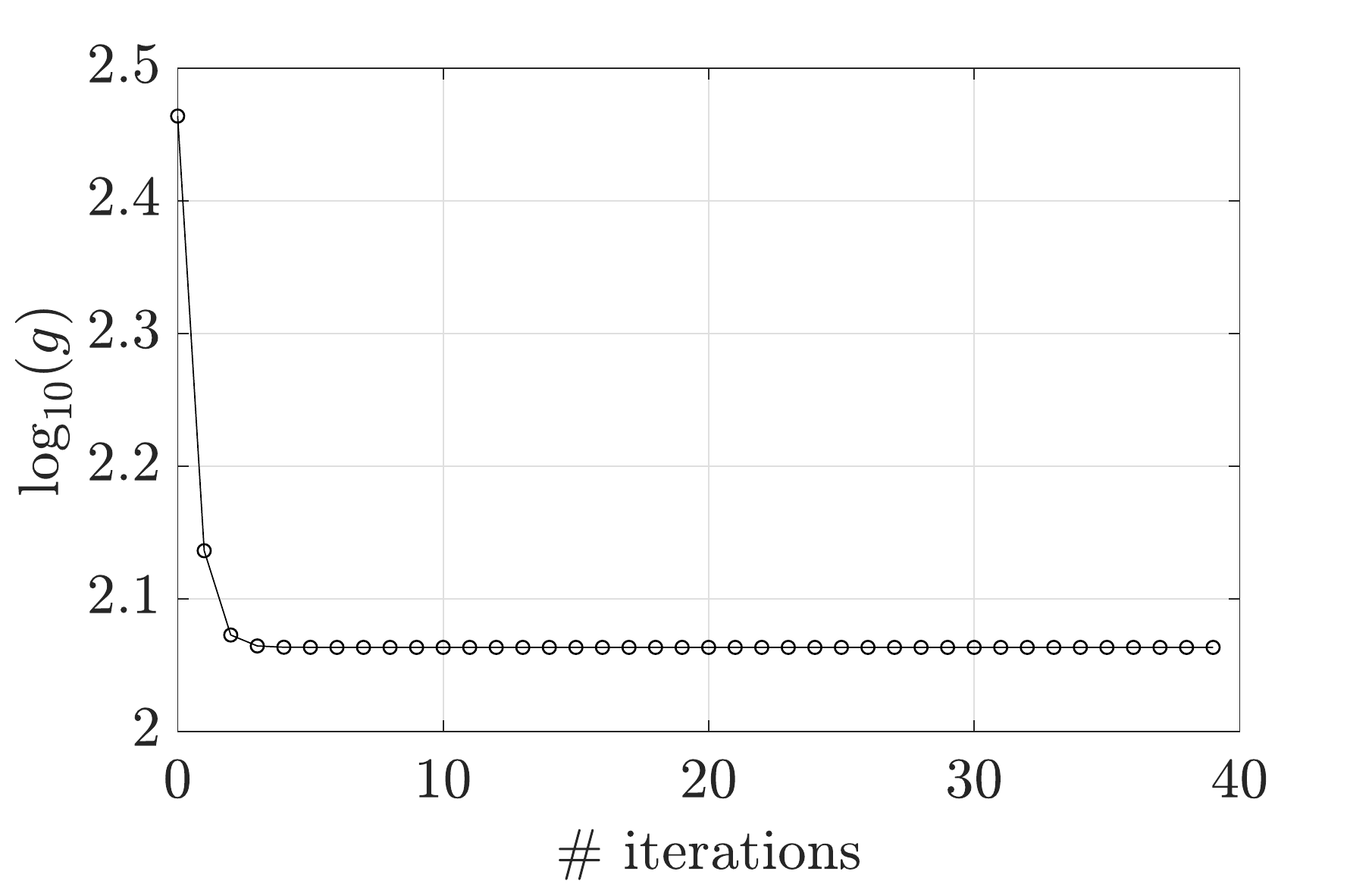}\label{fig:valid_2D_gCost}}
	\vspace{-1mm} \\
	\hspace{-0.2cm}
	\subfigure[Line tracking ($M=3$, $n=4$)]{\includegraphics[height=3cm,width=4.2cm]{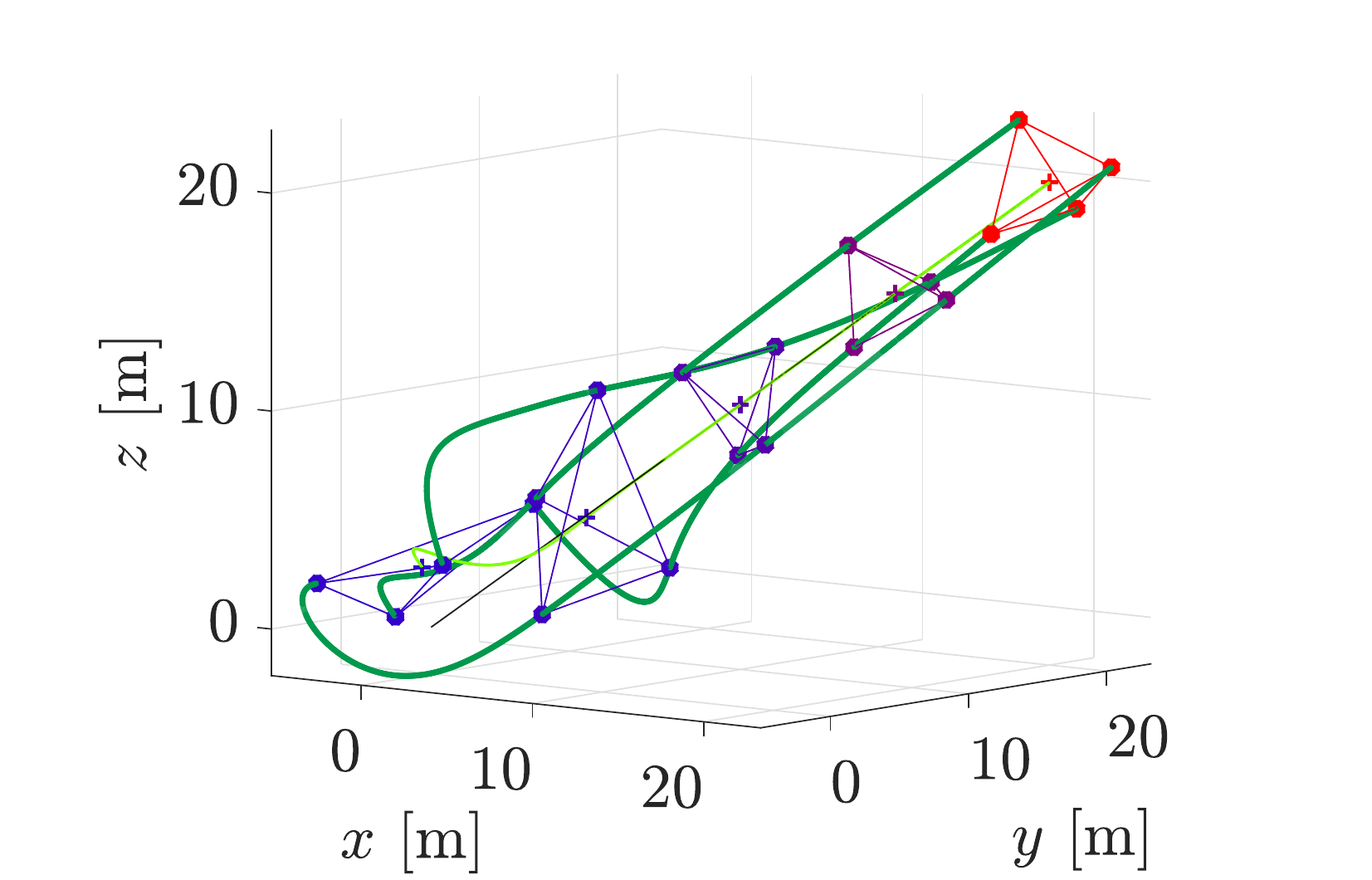}\label{fig:valid_3D_scene}}
	\hspace{-0.19cm} 
	\subfigure[Cost function in (c)]{\includegraphics[height=3cm,width=4.2cm]{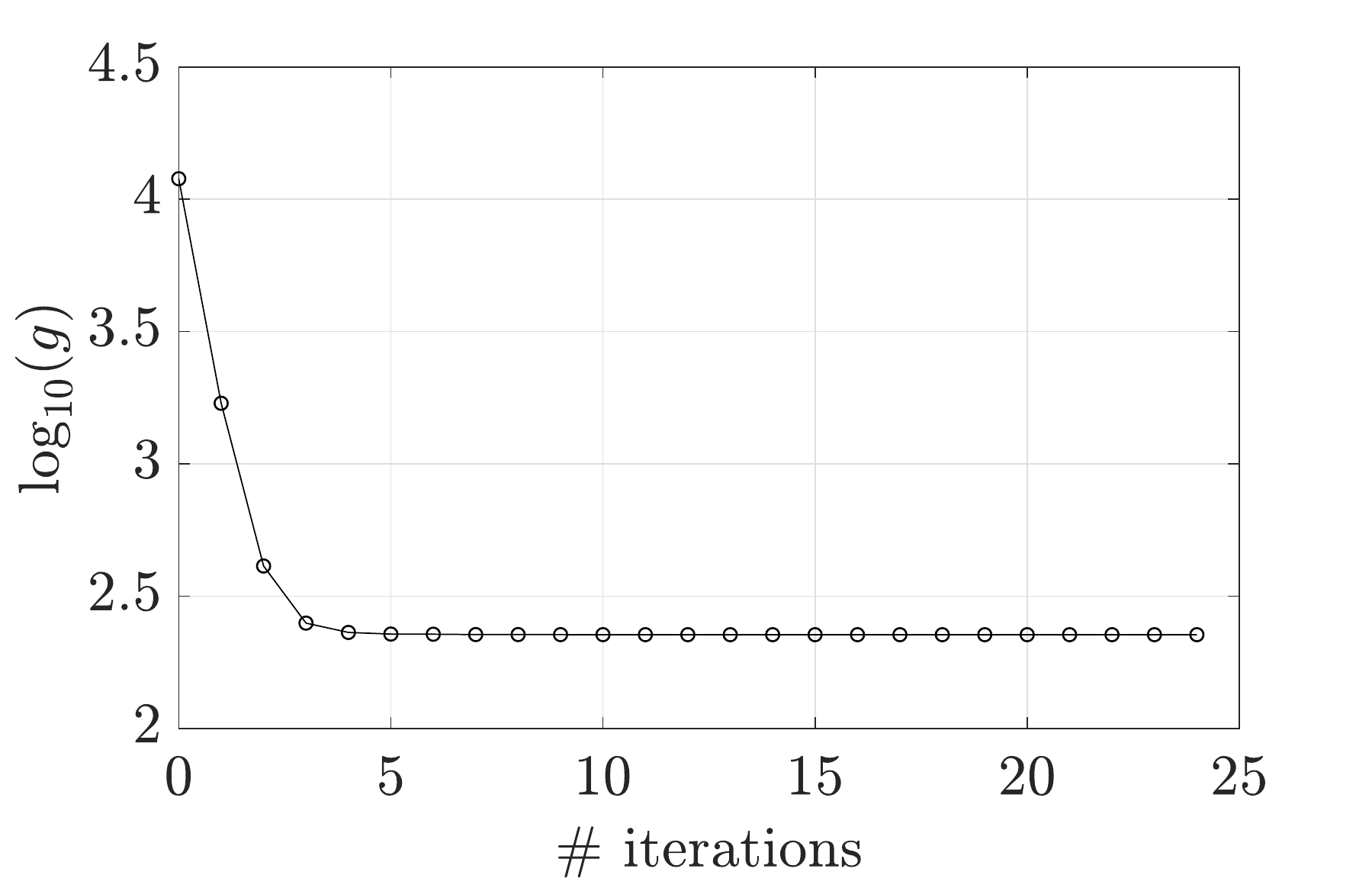}\label{fig:valid_3D_gCost}}
	\vspace{0mm}
	\caption{$\!\!\!\!\!\!$(a)-(b) 2D scenario: a straight line is tracked by $n=3$ agents that tend to reach the equilateral triangle configuration $(2,3)$. 
	$~~~~~~~~~~~~~~~~~~~~~~~~~~~~~~~~~~~~~$(c)-(d) 3D scenario: a straight line is tracked by $n=4$ agents that tend to reach the tetrahedron configuration $(3,4)$.}
	\label{fig:validation2D3D}
\end{figure}

In the two simulations depicted in Fig. \ref{fig:validation2D3D}, the tracking of a straight line at the constant velocity $\mathrm{v} = 1~\si{\meter \per \second}$ is required to be performed by two formations composed of $n=3$ and $n=4$ agents in a planar ($M=2$) and three-dimensional ($M=3$) scenarios respectively. Denoting with $\mathrm{vec}(\cdot)$ the vectorization operator\footnote{By definition, this operator stacks the vectors assigned to its argument, i.e. $\mathrm{vec}\left(\mathbf{w}_{1},...,\mathbf{w}_{\natural}\right) =\begin{bmatrix}
	\mathbf{w}_{1}^{\top} & \dots & \mathbf{w}_{\natural}^{\top}
	\end{bmatrix}^{\top}$.}, the initial conditions are set as
\begin{align*}
&\mathbf{p}(0) = \mathrm{vec}\left(\begin{bmatrix}
-2 \\ 1
\end{bmatrix},
\begin{bmatrix}
-3 \\ -1
\end{bmatrix},
\begin{bmatrix}
2 \\ -2
\end{bmatrix}\right)
\si{\meter}\\
&\dot{\mathbf{p}}(0) = \mathrm{vec}\left(\begin{bmatrix}
0 \\ -5
\end{bmatrix},
\begin{bmatrix}
0 \\ -5
\end{bmatrix},
\begin{bmatrix}
0 \\ -5
\end{bmatrix}\right)
\si{\meter \per \second}
\end{align*}
for the scenario in Fig. \ref{fig:valid_2D_scene};
\begin{align*}
&\mathbf{p}(0) = \mathrm{vec}\left(\begin{bmatrix}
-2 \\ 1 \\ 0
\end{bmatrix},
\begin{bmatrix}
-3 \\ -1 \\ 1
\end{bmatrix},
\begin{bmatrix}
2 \\ -2 \\ 2
\end{bmatrix},
\begin{bmatrix}
1 \\ 3 \\ 3
\end{bmatrix}\right)
\si{\meter}\\
&\dot{\mathbf{p}}(0) = \mathrm{vec}\left(\begin{bmatrix}
0 \\ -5 \\ 5
\end{bmatrix},
\begin{bmatrix}
0 \\ -5 \\ 0
\end{bmatrix},
\begin{bmatrix}
0 \\ -5 \\ 10
\end{bmatrix},
\begin{bmatrix}
0 \\ 0 \\ 0
\end{bmatrix}\right)
\si{\meter \per \second}
\end{align*}
for the scenario in Fig. \ref{fig:valid_3D_scene}. In addition, $q_{p} = 10 ~\si{\meter^{-2}}$, $q_{v} = 1 ~\si{\meter^{-2} \second^{2}}$ and $MaxIter = 50$ are assigned.

It is worth to note that the functional $g$ maintains strictly decreasing while the execution is running and both simulations correctly terminate before the maximum number of iterations is reached. Moreover, it can be numerically shown that instantaneous costs $l^{tr}$ in \eqref{eq:track_inst_cost} and $l_{d}^{fo}$ in \eqref{eq:form_inst_cost} tend to zero as the time $t$ goes to $T$, since $d$-configurations that satisfy all the constraints in \eqref{eq:constraints_d_equal} are attained and their centroids converge towards the desired trajectory.

\subsection{Equilibria of the final configurations}
\begin{figure}[h!]
	\centering
	\hspace{-0.0cm}
	\subfigure[Configuration $(2,5)$ ]{\includegraphics[height=3cm,width=4.2cm]{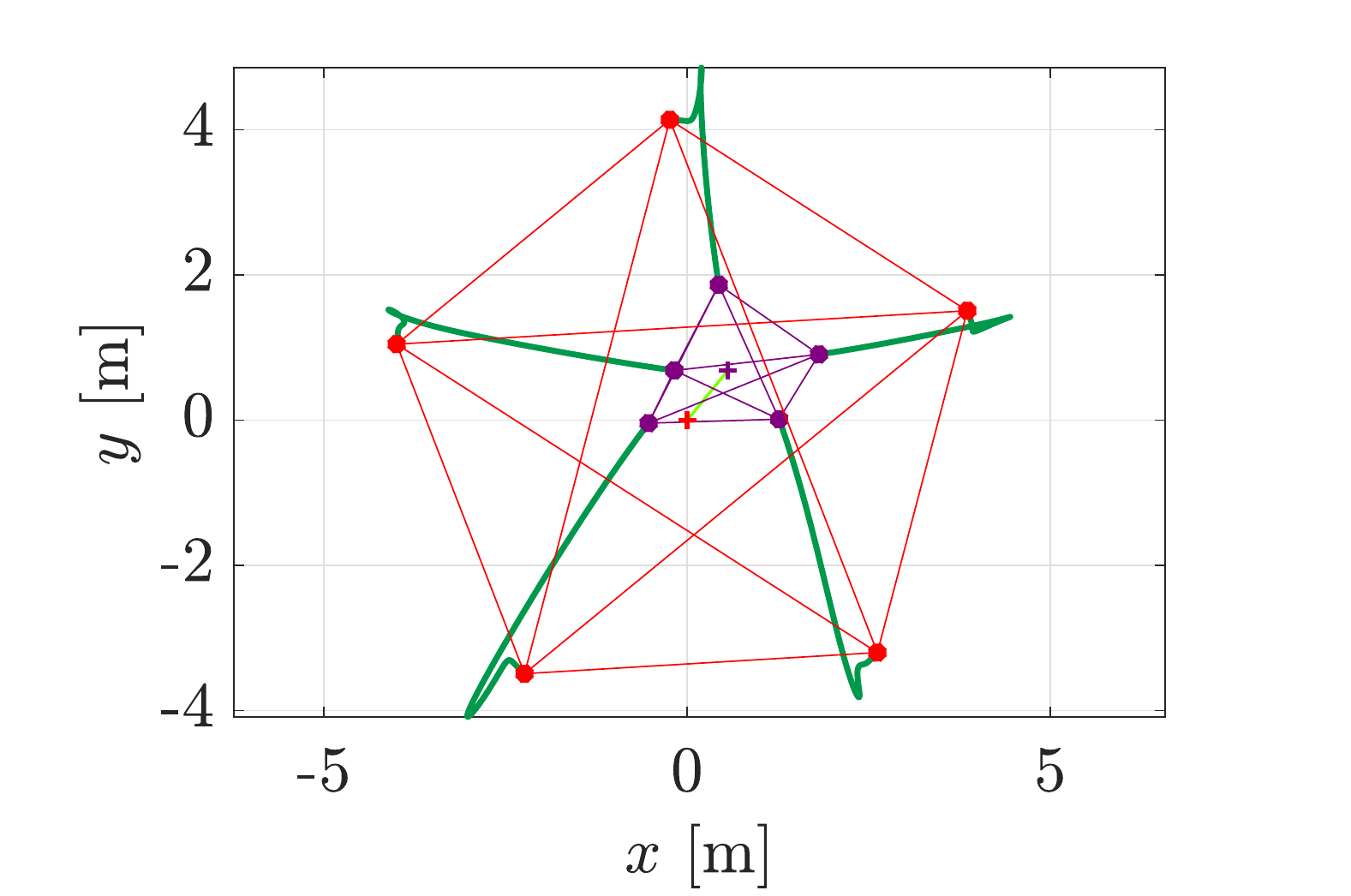}\label{fig:eq5_2D}}
	\hspace{-0.4cm}
	\subfigure[Configuration $(3,5)$]{\includegraphics[height=3cm,width=4.4cm]{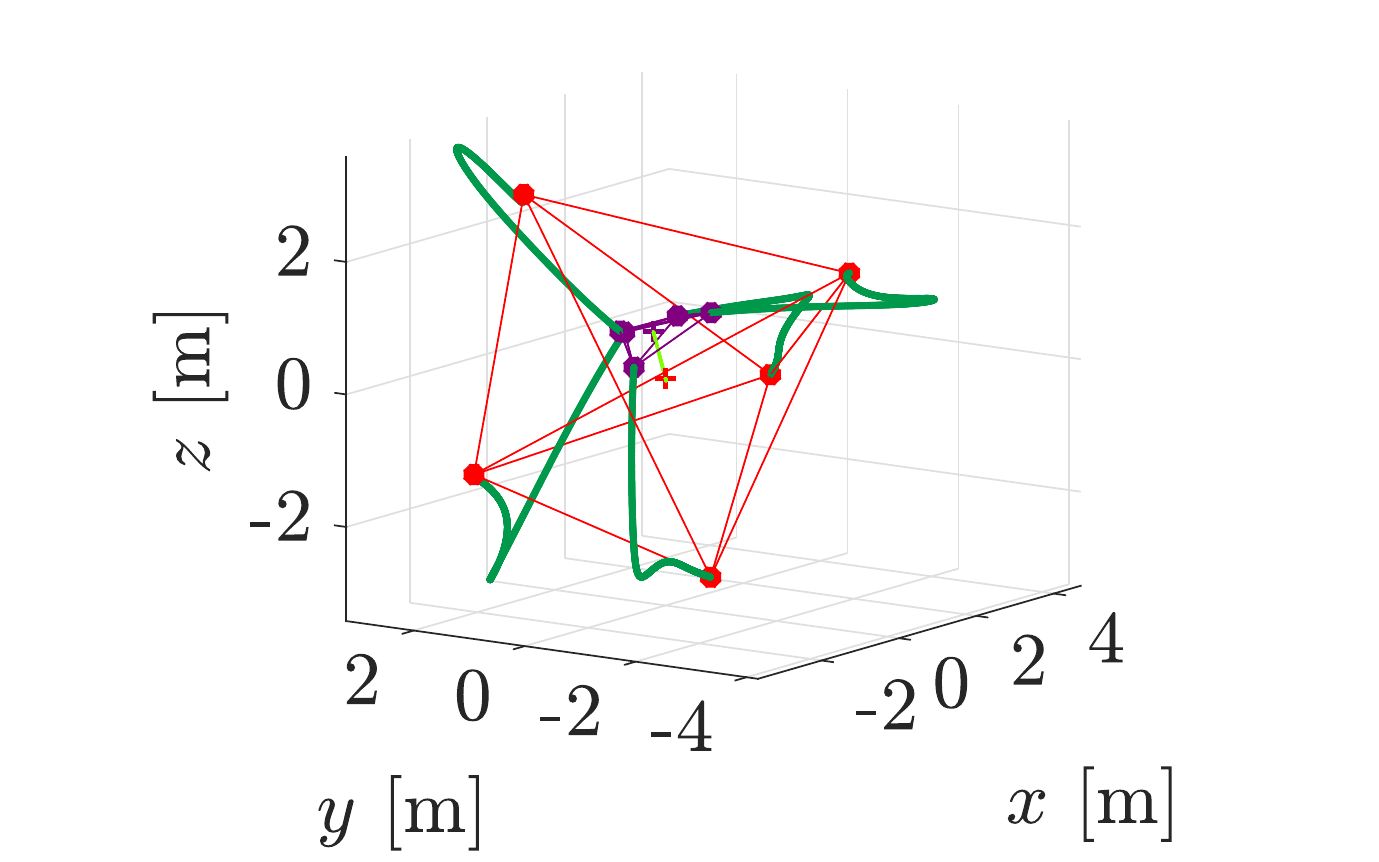}\label{fig:eq5_3D}}\\
	\hspace{-0.0cm}
	\subfigure[Configuration $(2,6)$]{\includegraphics[height=3cm,width=4.2cm]{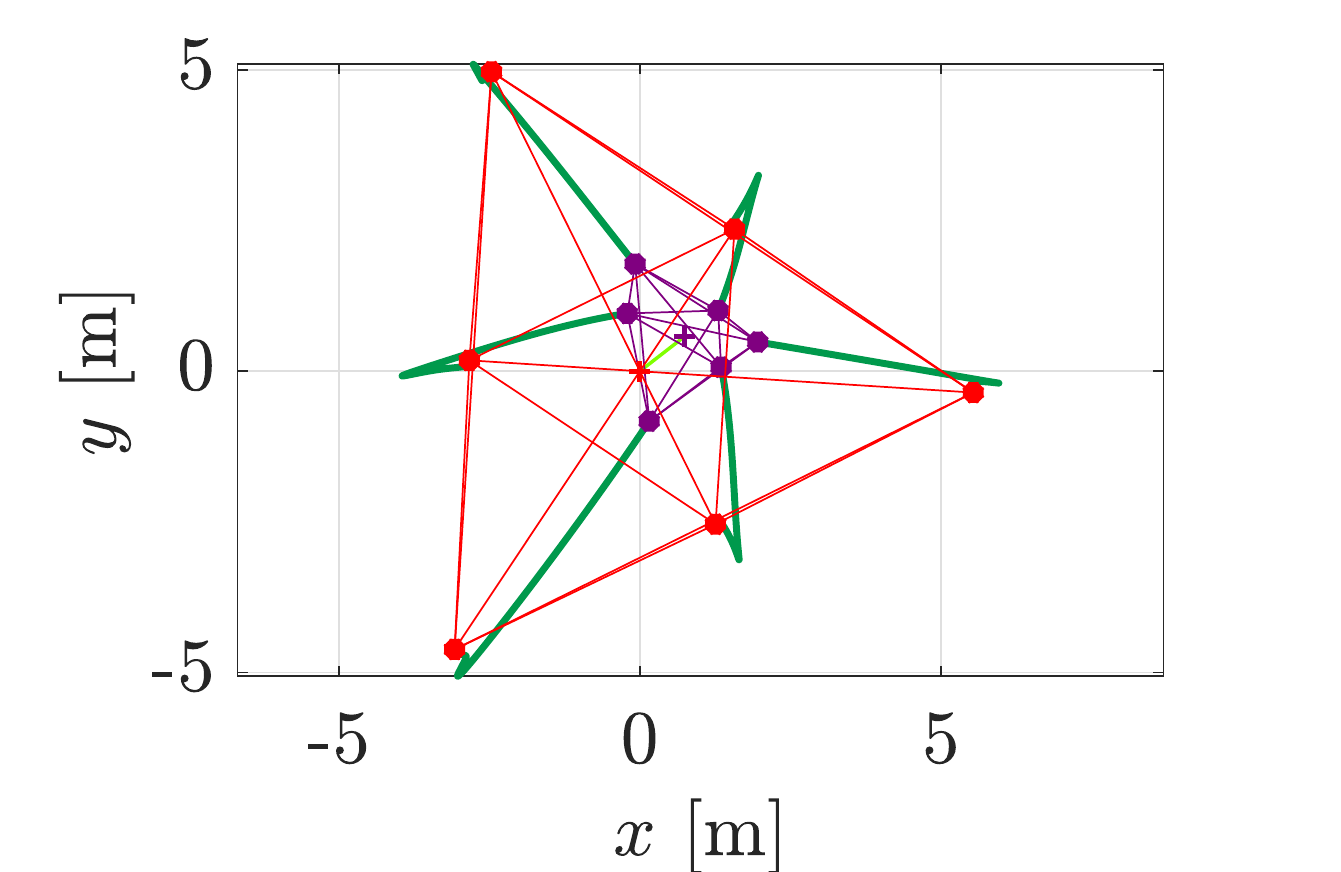}\label{fig:eq6_2D}}
	\hspace{-0.4cm}
	\subfigure[Configuration $(3,6)$]{\includegraphics[height=3cm,width=4.4cm]{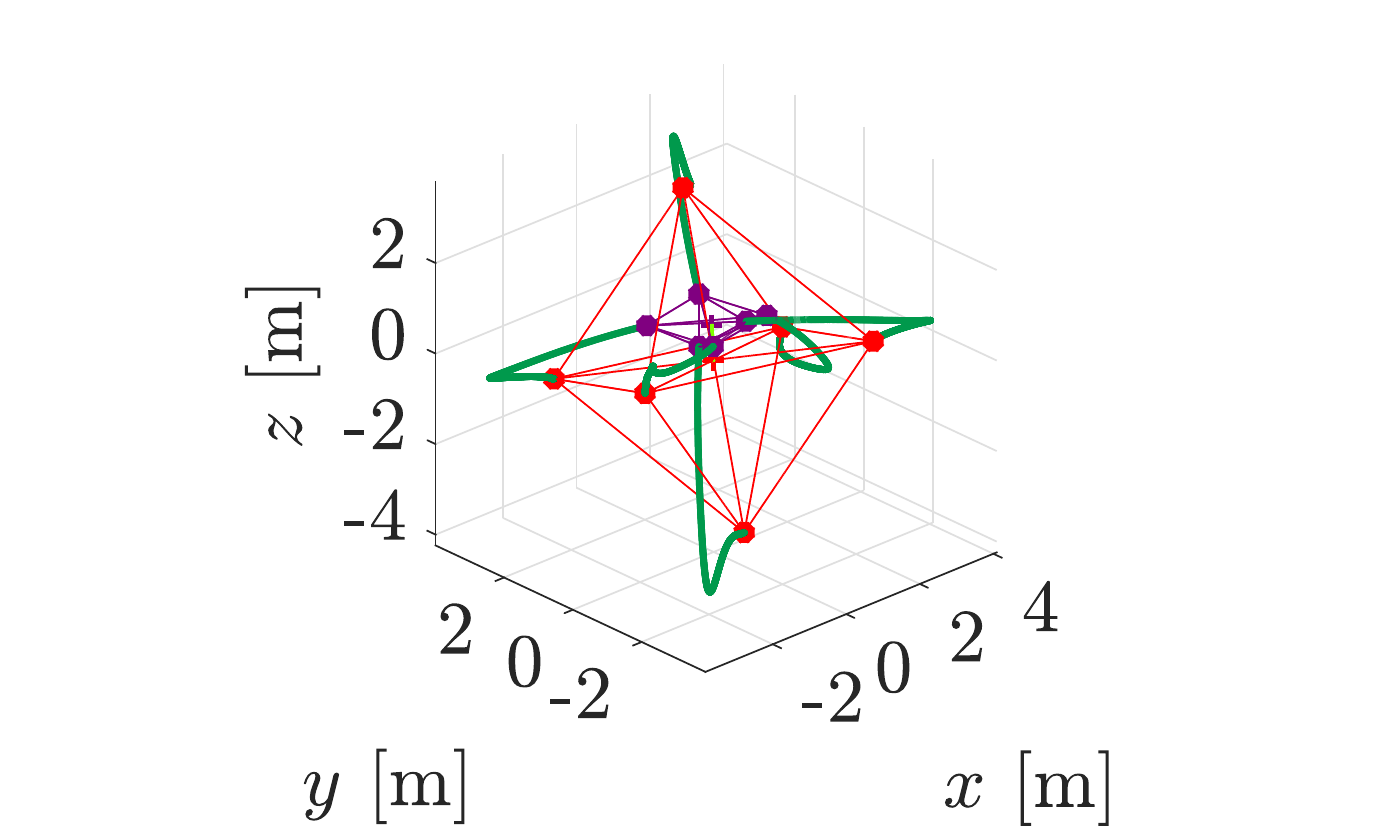}\label{fig:eq6_3D}}\\
	\hspace{-0.0cm}
	\subfigure[Configuration $(2,8)$]{\includegraphics[height=3cm,width=4.2cm, trim={0 3mm 0 0},clip]{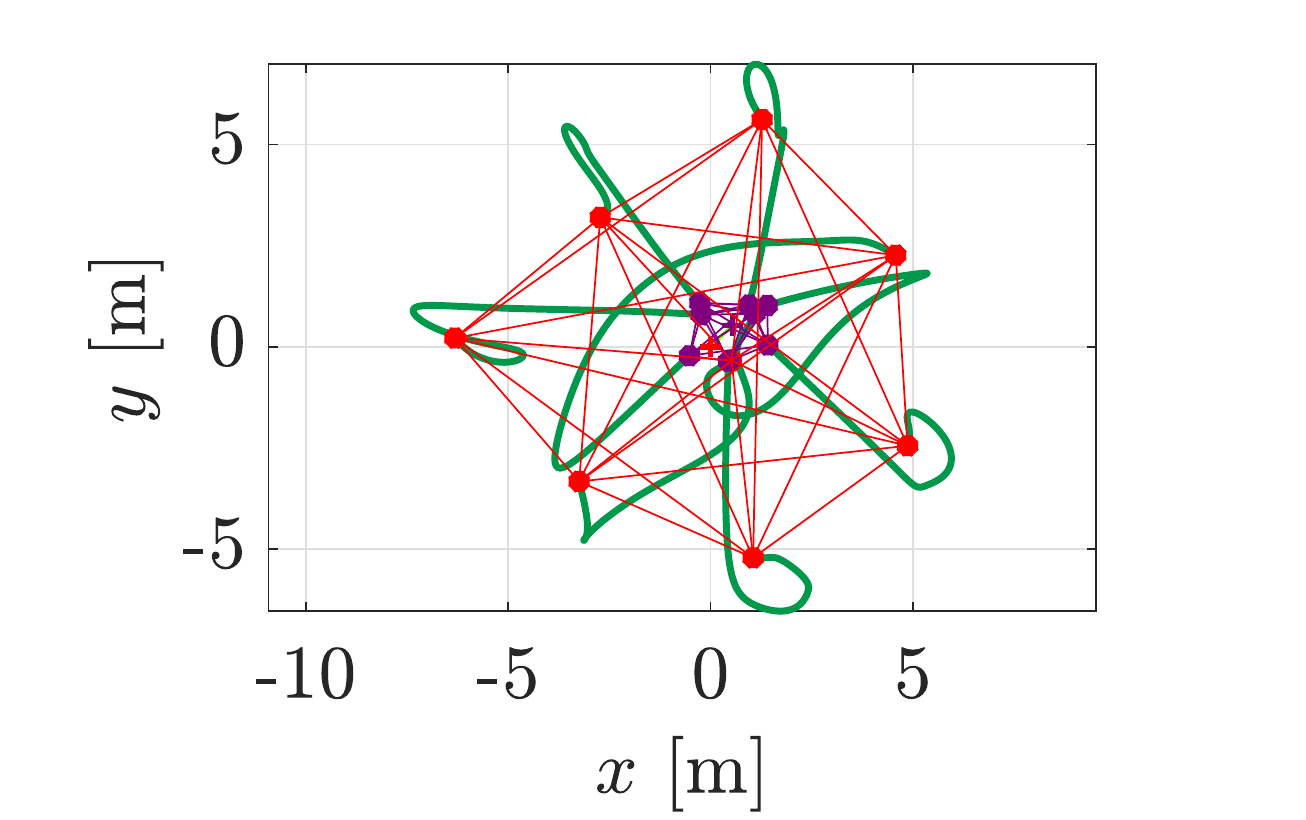}\label{fig:eq8_2D}}
	\hspace{-0.4cm}
	\subfigure[Configuration $(3,8)$]{\includegraphics[height=3cm,width=4.4cm]{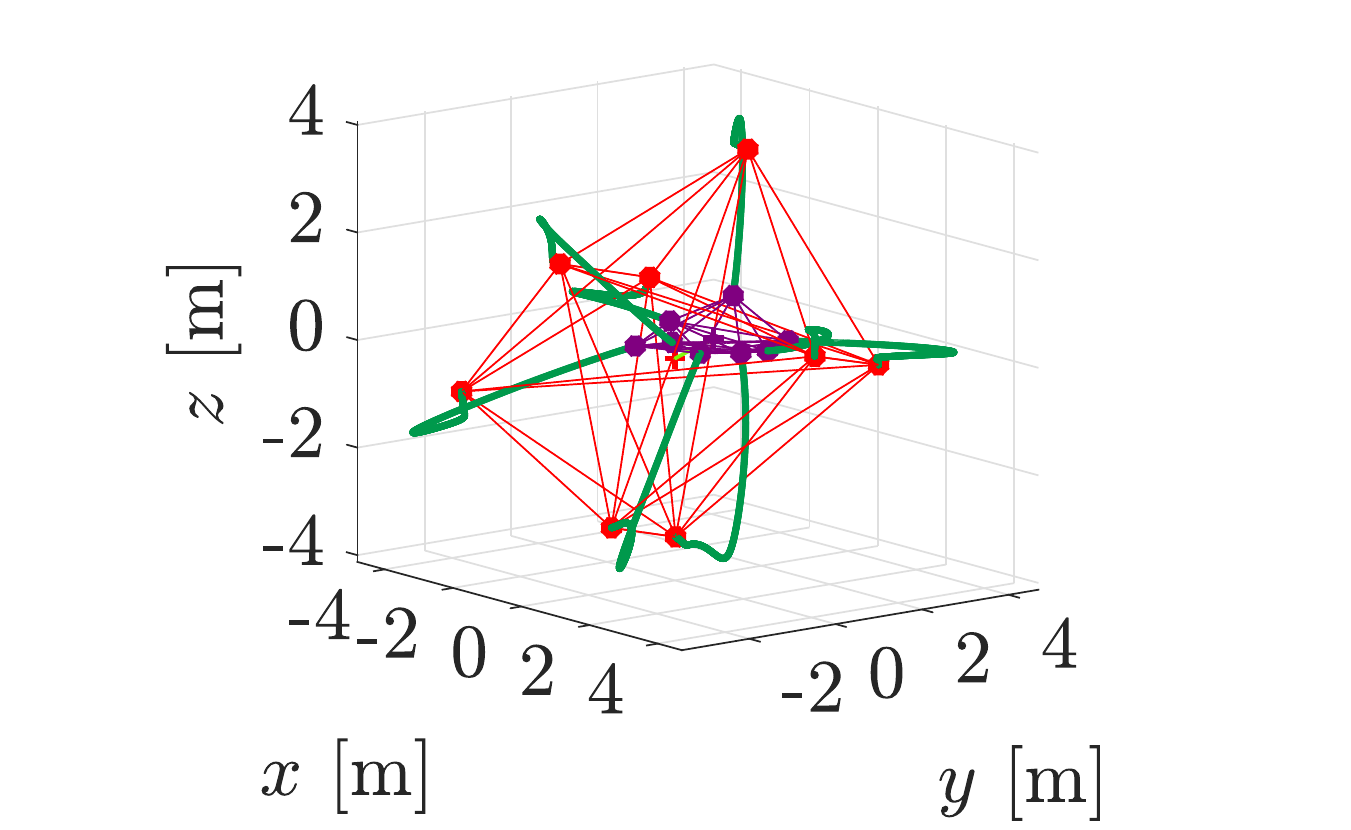}\label{fig:eq8_3D}}
	\caption{$\!\!\!\!$Left: equilibrium for final configurations achieved by $n=5,6,8$ robots in a 2D scenario. Agents converge toward shapes close to a regular pentagon~(a), a six-point truss~(c) and a eight-point truss~(e). 
	$~~~~~~~~~~~~~~~~~~~~~~~~~~$Right: equilibrium for final configurations achieved by $n=5,6,8$ robots in a 3D scenario. Agents converge toward shapes close to a double tetrahedron~(b), an octahedron~(d) and a uniform square antiprism~(f).}
	\label{fig:eq2D3D}
\end{figure}

In the group of simulations illustrated in Fig. \ref{fig:eq2D3D}, the equilibria of few particular formations are depicted: in this setup, agents are randomly deployed near the origin at the beginning
and then asymptotically driven towards it. Therefore,  $q_{p} = 10 ~\si{\meter^{-2}}$, $q_{v} = 1 ~\si{\meter^{-2} \second^{2}}$ and $MaxIter = 100$ are assigned.
Assuming that a constraint of set \eqref{eq:constraints_d_equal} is satisfied if \eqref{eq:constraints_d_equal_relaxation} holds true, it is important to note that, for some equilibria accomplished in these simulations, the maximum number of constraints in \eqref{eq:constraints_d_equal} is not successfully attained; however, most of the final configurations obtained possess similar shapes of actual $d$-configurations.

Let $\varphi_{c}$ be the ratio between the number of constrains satisfied in the set \eqref{eq:constraints_d_equal} for a specific solution and its cardinality. Table \ref{tab:distances_comp} shows a comparison between $\varphi_{c}$ and its maximum allowed value for simulations in Fig. \ref{fig:eq2D3D}.

\begin{center}
	\begin{table}[h!]
		\caption{Comparison between $\varphi_{c}$ (left) and $\max \varphi_{c}$ (right)}\label{tab:distances_comp}
		\begin{tabular}{|c|c|c|c||c|c|c|c|}
			\hline
			\multicolumn{4}{|c||}{Configurations achieved}                                                                         & \multicolumn{4}{c|}{$d$-configurations}                                                    \\ \hline
			\multicolumn{1}{|c|}{$M \diagdown n$} & \multicolumn{1}{c|}{$5$} & \multicolumn{1}{c|}{$6$} & \multicolumn{1}{c||}{$8$} & \multicolumn{1}{c|}{$M \diagdown n$} & \multicolumn{1}{c|}{$5$} & $6$ & \multicolumn{1}{c|}{$8$} \\ \hline
			$2$ & 5/10 & 9/15 & 12/28 & $2$ & 7/10 & 9/15 & 12/28 \\ \hline
			$3$ & 6/10 & 12/15 & 14/28 & $3$ & 9/10 & 12/15 & 16/28 \\ \hline
		\end{tabular}
	\end{table}
\end{center}

The fact that, in many cases, final configurations look like $d$-configurations might mean a good indication that a global minimum for $g$ is attained in each proposed example, since cost $l_{d}^{fo}$ takes into account the total sum of the potentials for each $(i,j)$ only. On the other hand, it is remarkably worth to recall that this term cannot be generally minimized just by satisfying the maximum number of constraints in \eqref{eq:constraints_d_equal}, as example $(2,5)$ in Fig. \ref{fig:eq5_2D} shows. Indeed, even though a five-point truss is the best shape to achieve, a regular pentagon seems to represent a lower minimum for $l_{d}^{fo}$, probably because the latter is more compact in terms of inter-agent distances.

\subsection{Complex trajectory tracking}\label{subs:Complex_trajectory_tracking}
\begin{figure}[t!]
	\centering
	\subfigure[Hyperbolic tangent tracking ($n=3$, $M=2$)]{\includegraphics[height=3.9cm,width=6.31cm]{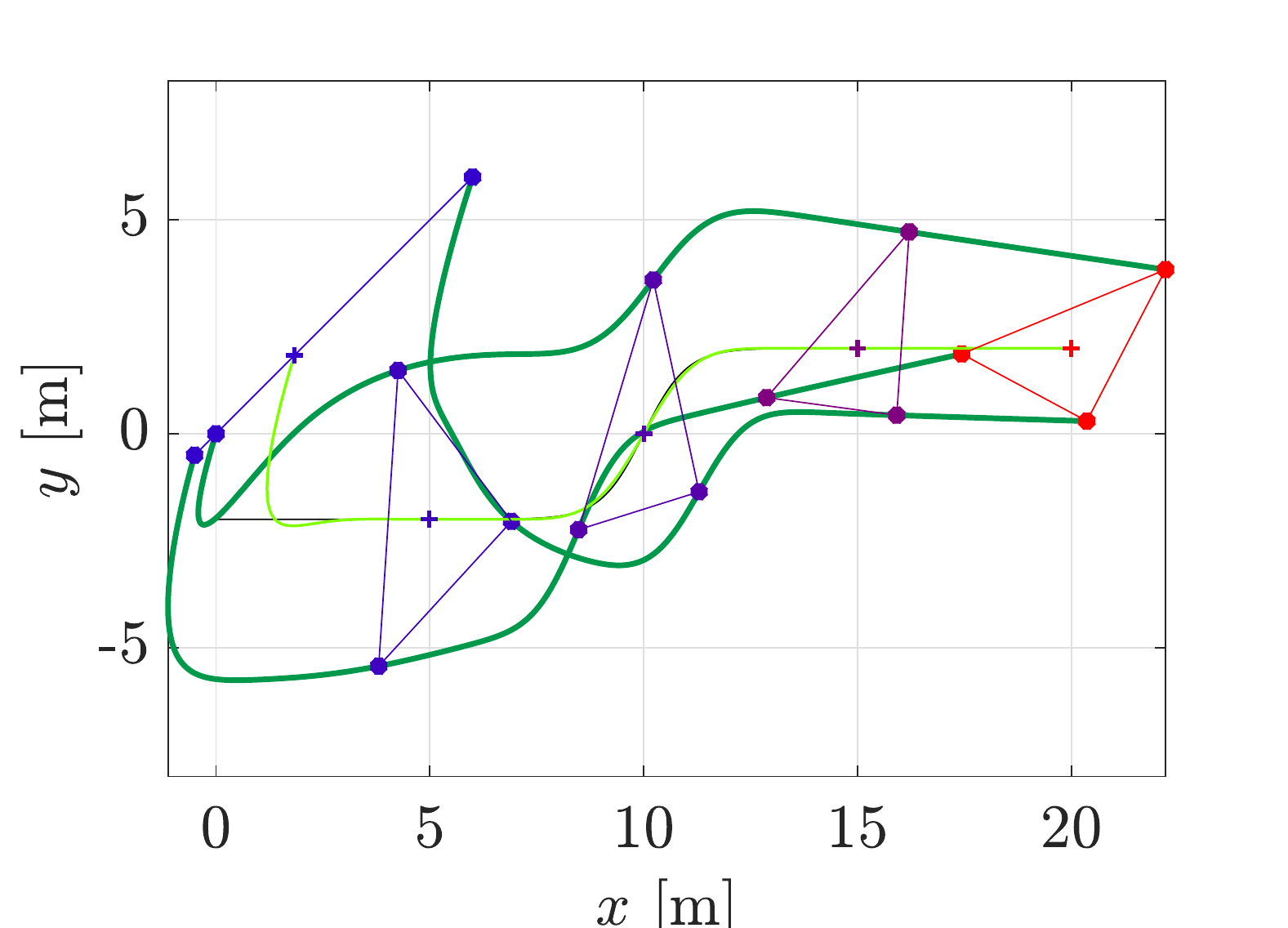}\label{fig:compl_traj_2D}}
	\vspace{-2mm}\\ 
	\subfigure[Helix curve tracking ($n=4$, $M=2$)]{\includegraphics[height=4.5cm,width=7.28cm]{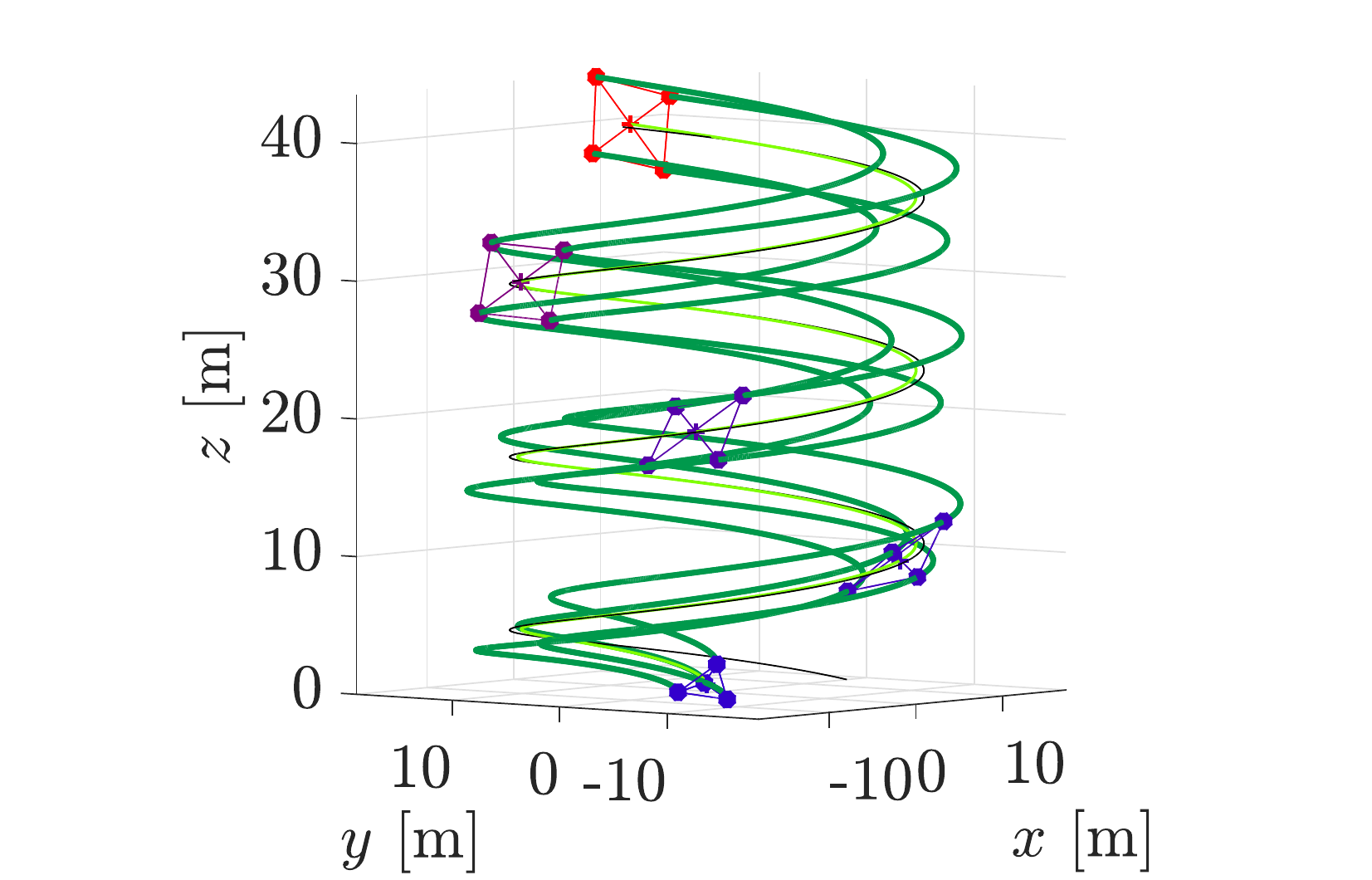}\label{fig:compl_traj_3D}}
	\vspace{3mm}
	\caption{(a)-(b): tracking of complex desired trajectories not involving $d$-configurations. In (a) and (b) a rectangular triangle and a planar square formations are achieved respectively.}
	\label{fig:compl_traj_2D3D}
\end{figure}

In this subsection, complex desired trajectories are required to be tracked by agents' barycenter while feasible constraints with the same form of \eqref{eq:dist_constraints} are imposed. In Fig. \ref{fig:compl_traj_2D}, $n=3$ agents follow a curve $\gamma_{2}(t)$ parametrized as
\begin{equation*}
\gamma_{2}(t) : \begin{cases}
x(t) = \mathrm{v}t\\
y(t) = \mathrm{r}\tanh(t-T/2)
\end{cases}, \quad t\in [0,T],
\end{equation*}
where $\mathrm{r} = 2~\si{\meter}$ and $\mathrm{v} = 1~\si{\meter \per \second}$. In this framework, robots are also required to gather in a rectangular-triangle formation characterized by desired inter-agent distances equal to $(3,4,5)~\si{\meter}$. Initial conditions for the system are set as follows:
\begin{align*}
&\mathbf{p}(0) = \mathrm{vec}\left(\begin{bmatrix}
-0.5 \\ -0.5
\end{bmatrix},
\begin{bmatrix}
0 \\ 0
\end{bmatrix},
\begin{bmatrix}
6 \\ 6
\end{bmatrix}\right)
\si{\meter}, \quad \dot{\mathbf{p}}(0) = \mathbf{0}_{N}.
\end{align*}
Whereas, in Fig. \ref{fig:compl_traj_3D}, $n=4$ agents follow a curve $\gamma_{3}(t)$ parametrized as
\begin{equation*}
\gamma_{3}(t) : \begin{cases}
x(t) = \mathrm{r}\cos(t)\\
y(t) = \mathrm{r}\sin(t)\\
z(t) = \mathrm{v} t
\end{cases}, \quad t\in [0,T],
\end{equation*}
where $\mathrm{r} = 15~\si{\meter}$ and $\mathrm{v} = 2~\si{\meter \per \second}$. In this framework, agents are also required to gather in a planar squared formation characterized by desired inter-agent distances equal to $5~\si{\meter}$ (sides) and $5\sqrt{2}~\si{\meter}$ (diagonals). Initial conditions for the system are set as follows:
\begin{align*}
&\mathbf{p}(0) = \mathrm{vec}\left(\begin{bmatrix}
-5 \\ -5 \\ 0
\end{bmatrix},
\begin{bmatrix}
0 \\ 0 \\ 2
\end{bmatrix},
\begin{bmatrix}
6 \\ 6 \\ 0
\end{bmatrix},
\begin{bmatrix}
-2 \\ 2 \\ 0
\end{bmatrix}\right)
\si{\meter}, \quad \dot{\mathbf{p}}(0) = \mathbf{0}_{N}.
\end{align*}
In both cases, $q_{p} = 10^{2} ~\si{\meter^{-2}}$, $q_{v} = 1 ~\si{\meter^{-2} \second^{2}}$ and $MaxIter = 50$ are assigned, since we have noticed that the value of $q_{p}$ visibly affects the tracking performances\footnote{Except for the transient phase, the instantaneous displacement $\left\|\mathbf{p}_{B}(\cdot)-\mathbf{p}_{B,des}(\cdot)\right\|$ significantly decreases when $q_{p}$ is increased.}, especially in the three-dimensional case.

\subsection{Invariant properties of the solution}
\begin{figure}[h!]
	\centering
	\subfigure[Movement in a $1$-dimensional subspace]{\includegraphics[height=3.9cm,width=6.31cm]{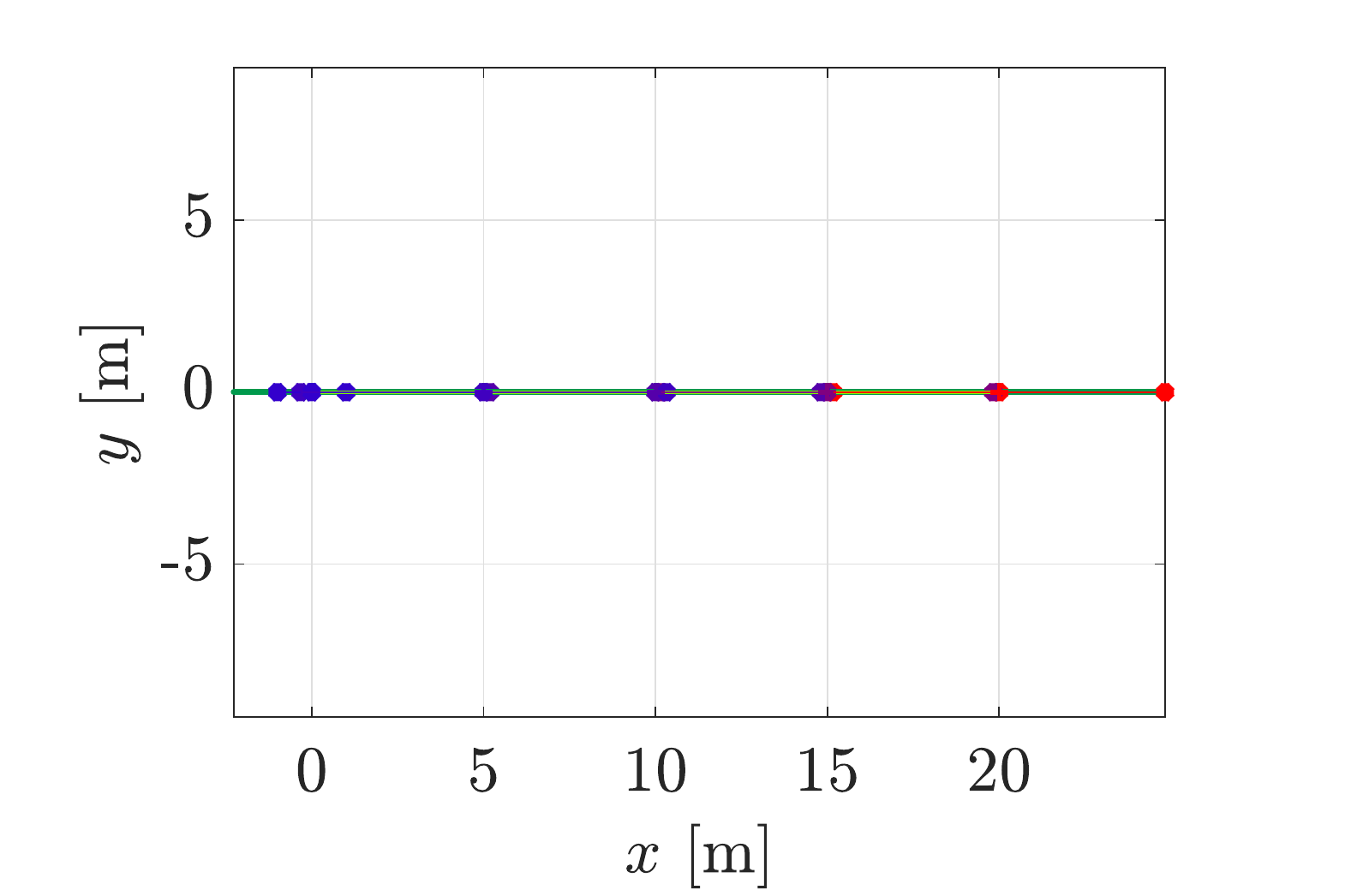}\label{fig:subs_2D}}
	\vspace{-2mm}\\
	\subfigure[Movement in a $2$-dimensional subspace]{\includegraphics[height=4.5cm,width=7.28cm]{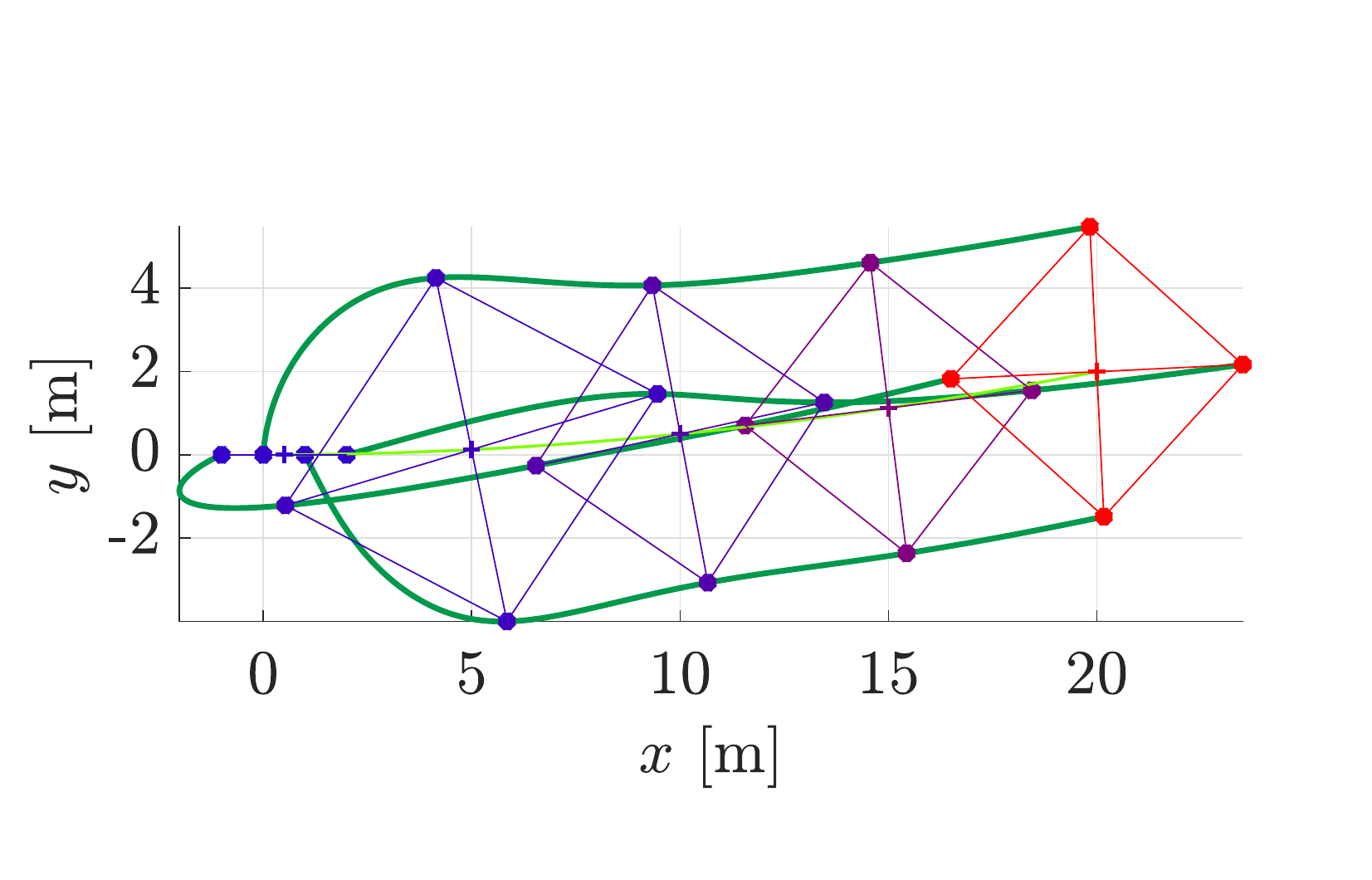}\label{fig:subs_3D}}
	\vspace{3mm}
	\caption{(a)-(b): initial conditions and desired trajectories belonging to a subspace $\mathcal{S}$ of inferior dimension seems to imply that the entire agents' movement remains within $\mathcal{S}$. (a) In a planar scenario, subspace $\mathcal{S}=\text{span}\{(x,0)\}$ represents a straight line corresponding to the $x$ axis. (b) In a three-dimensional scenario, subspace $\mathcal{S}=\text{span}\left\lbrace (x,0,0),\left(0,\dfrac{1}{200}x^{2},0\right)\right\rbrace$ represents a parabola laying on the $xy$ plane (contour for $z=0$).}
	\label{fig:subs_2D3D}
\end{figure}

One intriguing property appears during this study: initial conditions and desired trajectory could affect the evolution of the system in terms of invariant sets. More precisely, we can state the following
\begin{conj}
	\textit{Let $\mathcal{S}$ be a subspace of $\mathbb{R}^{M}$ such that $\mathbf{x}_{B,des}(\cdot)\in \mathcal{S}$ and $\mathbf{x}(0)\in \mathcal{S}$. By assuming all the hypotheses stated in Sec. \ref{sec:problem_setup}, the trajectory $\mathbf{x}(\cdot)$ computed by Alg. \ref{alg:PRONTO} entirely belongs to $\mathcal{S}$.}
\end{conj}
This conjecture is supported by the analysis provided in the forthcoming simulation, where $q_{p} = 10 ~\si{\meter^{-2}}$, $q_{v} = 1 ~\si{\meter^{-2} \second^{2}}$ and $MaxIter = 50$ are assigned and initial conditions for scenarios in Fig. \ref{fig:subs_2D} and Fig. \ref{fig:subs_3D} are set to be
\begin{align*}
&\mathbf{p}(0) = \mathrm{vec}\left(\begin{bmatrix}
-1 \\ 0
\end{bmatrix},
\begin{bmatrix}
0 \\ 0
\end{bmatrix},
\begin{bmatrix}
1 \\ 6
\end{bmatrix}\right)
\si{\meter}, \quad \dot{\mathbf{p}}(0) = \mathbf{0}_{N}
\end{align*}
and 
\begin{align*}
&\mathbf{p}(0) = \mathrm{vec}\left(\begin{bmatrix}
-1 \\ 0 \\ 0
\end{bmatrix},
\begin{bmatrix}
0 \\ 0 \\ 0
\end{bmatrix},
\begin{bmatrix}
1 \\ 0 \\ 0
\end{bmatrix},
\begin{bmatrix}
2 \\ 0 \\ 0
\end{bmatrix}\right)
\si{\meter}, \quad \dot{\mathbf{p}}(0) = \mathbf{0}_{N}
\end{align*}
respectively.

\section{Conclusions and future directions} \label{sec:conclusions}
We have formalized a simplified OIFT problem whose minimum-energy solution can be computed by a numerical tool called PRONTO. In order to achieve precise final configurations for the formations we want to drive, a meticulous use of potential functions have been made. In particular, simulations not only show the correctness of this approach but also give a fair description of its robustness when infeasible geometric constraints are imposed to the inter-agent distances. Numerical results also exhibit the versatility of this algorithm, since relatively complex desired trajectories can be tracked by tuning opportunely the weighting parameters contained in a suitable cost function that captures all the requirements. 

Possible future directions for this work might be represented by the application of PRONTO to a real OIFT problem with a nonlinear dynamics or the extension to a time-varying formation framework as well as a decentralized paradigm envisaging clusters of mobile robots with relevant communication constraints on the information sharing. Further inspections on the asymptotic stability for a system of agents steered by this optimization approach may likewise depict a fascinating research in this area.




\section*{Acknowledgments}
The authors would like to thank the association \textquotedblleft Fondazione Ing. Aldo Gini\textquotedblright$~$that supported this research and the University of Colorado at Boulder for its hospitality.





\bibliographystyle{IEEEtran}
\bibliography{IEEEfull,biblio}

\end{document}